\documentclass{aa}  
\usepackage{graphics}
\begin{document}
%
\thesaurus{ 2        
(
11.01.2  
11.05.2  
11.17.3  
12.04.2  
13.25.2  
13.25.3  
)}
\title{Soft X-ray AGN Luminosity Function from {\it ROSAT} Surveys}
\subtitle{I. Cosmological Evolution and
   Contribution to the Soft X-ray Background}
\author{Takamitsu Miyaji    \inst{1,2}
\thanks{Present address: Department of Physics, Carnegie Mellon
      University, Pittsburgh, PA15213-3890}
\and    G\"unther Hasinger  \inst{2}
\and    Maarten Schmidt     \inst{3}
}

\offprints{T. Miyaji (miyaji@xray.mpe.mpg.de)}

\institute{
Max-Planck-Inst. f\"ur Extraterrestrische Physik, Postf. 1603,
D--85740, Garching, Germany (miyaji@xray.mpe.mpg.de)
\and 
Astrophysikalisches Institut Potsdam, An der Sternwarte 16, D--14482
Potsdam, Germany (ghasinger@aip.de)
\and 
California Institute of Technology, Pasadena, CA 91125, USA  
(mxs@deimos.caltech.edu)
}

\date{Received date; accepted date}

\titlerunning{{\it ROSAT} AGN Luminosity Function} 
\authorrunning{Miyaji, Hasinger, Schmidt}

\maketitle
\begin{abstract}
 We investigate the evolution of the 0.5-2 keV soft X-ray luminosity 
function (SXLF) of active galactic nuclei (AGN) using results 
from {\it ROSAT} surveys of various depth. The large dynamic
range of the combined sample, from shallow large-area
{\it ROSAT} All-Sky Survey (RASS)-based samples to the deepest
pointed observation on the Lockman Hole, enabled us to trace the 
behavior of the SXLF. The combined sample includes about 690 AGNs. 
As previously found, the SXLF evolves rapidly as a function of 
redshift up to $z\sim 1.5$ and is consistent with remaining constant 
beyond this redshift.

 We have tried to find a simple analytical description of the
SXLF in the overall redshift and luminosity range, using  
Maximum-Likelihood fits
and Kolgomorov-Smirnov tests. We found that a form of the 
Luminosity-Dependent Density Evolution (LDDE), rather than 
the classical Pure Luminosity Evolution (PLE) or the 
Pure Density Evolution (PDE) models, gives an excellent 
fit to the data.  Extrapolating one form of the LDDE model 
(LDDE1) explains $\approx 60\%$ of the estimated soft extragalactic 
Cosmic X-ray Background (CXRB). We have also found another representation 
(LDDE2), which produces $\approx 90\%$ of the CXRB and 
still gives an excellent fit to the sample AGNs. These two
versions of the LDDE models can be considered two extremes of 
the possible extrapolations of the SXLF below the flux limit
of the survey.
    
 We have also investigated the evolution of the number density
of luminous QSOs with ${\rm Log}\,L_{\rm x}>44.5$ 
$[h_{\rm 50} {\rm erg\,s^{-1}}]$, where the evolution can 
be traced up to the high redshift. We have compared the
results with similar quantities in optically- and radio-selected
luminous QSOs. Unlike these QSOs,  evolution of the {\it
ROSAT}-selected QSOs do not show evidence for the decrease
of the number density in $z\ga 3$. The statistical significance
of the difference is, however, marginal.
 
\keywords{Galaxies: active -- Galaxies: evolution -- 
{\itshape (Galaxies:)} quasars: general --
{\itshape (Cosmology:)} diffuse radiation --  
X-rays:galaxies -- X-rays:general}
\end{abstract}

\section{Introduction}
\label{sec:intr}

 The AGN/QSO luminosity function and its evolution with 
cosmic time are key observational quantities on understanding
the origin of and accretion history onto supermassive balckholes,
which are now believed to occupy the centers of most galaxies. 
Since X-ray emission is one of the prominent characters of 
the AGN activity, X-ray  surveys are effective means of sampling 
AGNs for the luminosity function and evolution studies. 
The {\it R\"ontgen satellite} ({\it ROSAT}), with its unprecedented imaging 
capabilities, provided us with soft X-ray surveys with various 
depths, ranging from the {\it ROSAT} All-Sky Survey (RASS) 
to the {\it ROSAT} Deep Survey (RDS) on the Lockman Hole (Hasinger
et al. \cite{rds1}). Various optical identification programs 
of the survey fields have been conducted and the combination
of these now enabled us to construct the soft X-ray luminosity
function (SXLF) as a function of redshift. 

The evolution of SXLF
has already been seen in the Extended Medium Sensitivity Survey 
(EMSS) AGNs (Maccacaro et al. \cite{emss1}; Della Ceca et
al. \cite{emss2}) for high-luminosity AGNs. Combining results
from deep {\it ROSAT} PSPC surveys and the EMSS has extended 
the sample into the higher-redshift lower-luminosity regime, providing
much wider baseline to explore the evolution properties 
(e.g. Boyle et al. 1994; Jones et al. 1996; Page et al. 1996). 
All of these were characterized by a pure luminosity
evolution model (PLE) with approximately $\propto (1+z)^3$ 
up to $z\approx 2$, and consistent with no evolution beyond 
that point. Using a larger {\it ROSAT} sample, 
Page et al. (1997) found that PLE underpredicts the number of 
high-redhsift low luminosity AGNs for $q_0=0.5$.
Simple extrapolations of any of the 
PLE expressions only explain $\sim 30-50\%$ of the soft X-ray 
Background (0.5-2 keV) by AGN.
  
{Because of the relatively large PSF of the {\it ROSAT} PSPC, the
identifications of the deepest {\it ROSAT} PSPC surveys are sometimes 
ambiguous and misidentifications can occur.  
Based on results of the optical followup studies of {\it ROSAT} PSPC
surveys, a number of groups, including the Deep {\it ROSAT} Survey
(DRS; Griffiths et al. \cite{gri96}) and UK Deep Survey
(UKD;  McHardy et al. \cite{ukd}) report a population of X-ray 
sources called ``Narrow Emission-line Galaxies'' (NELG) at faint
fluxes. On the other hand, faint X-ray sources
found in the {\it ROSAT} Deep Survey  
on the Lockman Hole (RDS-LH), which have accurate source positions 
from 1 million seconds of {\it ROSAT} 
HRI data, are still predominantly AGNs down to the
faintest fluxe in the survey (Schmidt et al. \cite{rds2};
Hasinger et al. \cite{has99}).
Some of these have optical spectra which apparently show only
narrow-lines but have other signs of an AGN activity and might
have been classified as ``NELGs'' at the criteria of other
groups. On the other hand,  Lehmann et al. (\cite{rds3}) have 
compared redshift distributions
of the RDS X-ray AGNs, UKD X-ray sources, non X-ray emitting
(at the RDS-LH limit) field galaxies showing narrow 
emission-lines. They found that the redshift distribution of 
UKD X-ray sources has a significant excess over that of the 
RDS-LH sources at $z<0.4$. This excess was dominated by ``NELGs'', 
whose redshift distribution was similar to that of non X-ray source narrow
emission-line field galaxies. This shows that a significant fraction 
of ``NELGs'' are likely to be misidentifications by chance
coincidences. This observation seems to contradict with estimations
of the relatively low probabilities of such chance coincidences by the
DRS and UKD groups. A more detailed comparison is urgently needed.} 
Misidentifications affect SXLF estimates in two ways,
i.e., by putting a wrong object into the sample and by missing the 
true identifications. Thus it is important to have a high spatial
resolution image to obtain unambiguous identifications, especially
in the faintest regime. 
   
 In this study, we investigate the {global} behavior of the soft X-ray 
luminosity function (SXLF) of AGNs from a combined
sample of various {\it ROSAT} surveys. We use 
the term ``AGN'' for both Seyfert galaxies, including 
type 1's and type 2's, and QSOs. Preliminary work, using
earlier versions of the combined sample, have been reported in 
Hasinger (\cite{has_xs}) and 
Miyaji et al. (\cite{m99a}) (hereafter M99a), while in this work,
we have made a more extensive analysis with updated 
{\it ROSAT} Bright Survey (RBS) and {\it ROSAT} Deep Survey (RDS)
catalogs including new identifcations from observations 
made in the winter{-spring season of 1999}. 
{In this paper, we put emphasis on the expressions 
representing the global behavior of the SXLF. 
Presenting separate expressions in several redshift intervals,
giving more accurate representation of the data
in the redshift ranges of interest
will be a topic of a future paper (Miyaji et al. in preparation, 
paper II). In paper II, we will also present
tables of full numerical values of the binned SXLF.}      
  
We use a Hubble constant $H_0=50\,h_{50}\,$ 
$[{\rm km\,s^{-1}\,Mpc^{-1}}]$.
The $h_{50}$ dependences are explicitly stated. We calculate
the results with common sets of cosmological parameters: 
$(\Omega_{\rm m},\Omega_\Lambda) =$(1.0,0.0) and (0.3,0.0). 
For some important parameterized expressions, we also show 
the results for $(\Omega_{\rm m},\Omega_\Lambda) =$ (0.3,0.7).

\section{The {\it ROSAT} Surveys used in the analysis}
\label{sec:data}

 We have used soft X-ray sources identified with
AGNs with redshift information from  a combination of {\it ROSAT} 
surveys in various depths/areas from a number of already
published and unpublished sources. In order to avoid the possible
bias from the large-scale overdensity and the distortion
of the redshift-distance relation based on bulk-flows in the 
nearby universe  (e.g. Tully \& Shaya \cite{tusha}), we have
excluded objects within $z<0.015$ from the analysis.

 The surveys we have used are summarized in Table ~\ref{tab:surv}. 
Two optical followup programs from the 
{\it ROSAT} All-Sky Survey (RASS) (Voges \cite{rass}), a serendipitous 
survey of the {\it ROSAT} PSPC pointed 
observations (RIXOS), and a number of deep pointings specifically aimed 
for deep surveys. Here we describe the AGN sample from each survey.

 All surveys, except for a part of the 
the Lockman Hole, are based on the {\it ROSAT} PSPC count rates
in the pulse-invariant (PI) channel range corresponding to 
0.5 - 2 keV. For most sources in the Lockman Hole, we have used 
the deeper HRI count rates (see below) with no spectral resolution and
sensitive to the 0.1 - 2 keV. 

 In order to convert the countrate to flux, we have to assume
a spectrum. Hasinger et al. (\cite{has93})
obtained the value of $\Gamma=1.96\pm .11$ for the average
spectral photon index in the Lockman Hole. Other works (Romero-Colenero et al. 
\cite{rom}; Almaini et al. \cite{alm}) also found similar
spectral index for AGNs, but a harder index $\Gamma\approx 1.5$
for the ``NELGs''. The same class as a part of the 
population they have classified as NELGs may fall into our sample.
In any case, the {\it ROSAT} countrate 
to the unabsorbed 0.5-2 keV flux conversion has been made assuming a 
power-law with a photon index of $\Gamma = 2.0$ and 
corrected for the Galactic absorption. In effect, the Galactic
column density changes the response curves for the flux-to-countrate 
conversion. However, the extragalactic surveys are mainly concentrated
on the part of the sky where the Galactic absorption is low.
Typical values are $(0.5-1)\times 10^{20}$ ${\rm [cm^{-2}]}$ for the
deep surveys and a maximum of $16\times 10^{20}$ ${\rm [cm^{-2}]}$ for
a small portion of the sky covered by the RBS. Within this range,  
the conversion between the $S_{\rm x}$ (here and hereafter, 
$S_{\rm x}$ represents the 0.5-2 keV flux
and $S_{\rm x14}$ is the same quantity measured
in units of $10^{-14} [{\rm erg\,s^{-1}\,cm^{-2}}]$)
the {\rm ROSAT} PSPC countrate (in the corresponding channel range)
only weakly dependent on the spectral shape and varies  
by about $\pm 3$\% for spectral indices $\Gamma=2.0\pm 0.7$. 
We discuss the conversion for the HRI case in Sect. \ref{sec:lhsamp}. 

 For the computation of the SXLF, it is important to define the
available survey area as a function of limiting flux. 
In case there is incompleteness in the 
spectroscopic identifications, we have made the usual 
assumption that the redshift/classification distribution of these 
unidentified sources is the same as the identified sources
at similar fluxes. This can be attained by defining
the 'effective' survey area as the geometrical survey area 
multiplied by the completeness of the identifications.
{This assumption is not correct when
the source is unidentified due to non-random causes, e.g.,
no prominent emission lines in the observed spectrum.}
However, given the high completeness
of the samples used in our analysis, this does not
affect the results significantly, {except for the
faintest end of RDS-LH. We discuss the 
effects of the incompleteness at this faint end in Sect. 
\ref{sec:dmcom}.}

Below we summarize our sample selection and completeness
for each survey.  

\begin{table}
\caption[]{{\it ROSAT} Surveys used in the Analysis}
\begin{center}
\begin{tabular}{lccc}
\hline\hline
 Survey$^{\rm a}$ & $S^{\rm lim}_{\rm x14}$ & Area & No. of$^{\rm b}$  \\
       & $[{\rm erg\,s^{-1}\,cm^{-2}}]$ & $[{\rm deg}^2]$ & AGNs  \\
\hline
 RBS   & $\approx 250$ & $2.0 \times 10^{4}$ & 216 \\
 SA-N  & $\approx 13$ &   $685.$  & 130 \\
 RIXOS & $3.0$  &  $15.$   & 205 \\    
 NEP   & $1.0$  & $0.21$  & 13  \\
 UKD   & $0.5$  & $0.16$  & 29  \\
 RDS-Marano & $0.5$ & $0.20$  & 30  \\
 RDS-LH & $0.17-0.9$ & $0.30$ & 68  \\
\hline
\end{tabular}
\end{center}
\label{tab:surv}
$^{\rm a}$ Abbreviations -- RBS: The {\it ROSAT} Bright Survey, 
SA-N: The Selected Area-North, RIXOS: The {\it ROSAT} International 
X-ray Optical Survey, NEP: The North Ecliptic Pole
UKD: The UK Deep Survey, 
RDS-Marano: The {\it ROSAT} Deep Survey -- Marano field,  
RDS-LH: The {\it ROSAT} Deep 
Survey -- Lockman Hole. See text for references.
$^{\rm b}$ Excluding AGNs with $z<0.015$. 
\end{table}

\begin{figure}
\resizebox{\hsize}{!}{\includegraphics{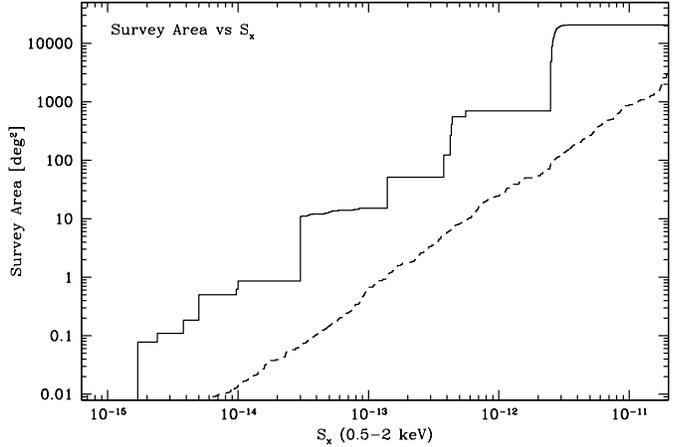}}
\caption[]{The survey area of the combined sample
are plotted as a function of the limiting 0.5-2 keV
flux limit (solid line). For reference, $[N(>S)]^{-1}$
for all the X-ray sources is overplotted (dashed-line).}
\label{fig:area}
\end{figure}

\subsection{The {\it ROSAT} Bright Survey (RBS)}

 The RBS program aims for a complete identification of the
$\sim$ 2000 brightest sources in the {\it ROSAT} All-Sky Survey
(RASS) for $|b|>30\deg$ (Fischer et al. \cite{rbs0};
Schwope et al. in preparation 
) measured in the
entire {\it ROSAT} band (0.1-2.4 keV). For our purposes,
we have extracted a subset selected by the {\it ROSAT}
Hard band (0.5-2 keV) countrate of $CR_{\rm hard}\geq 0.2$ 
$[{\rm cts\;s^{-1}}]$, which makes a complete hard 
countrate-limited sample. Five sources in this subsample
have further been identified as AGNs since M99a and included
in the analysis. This subsample now has  been 
completely identified. 

 Since the absorption in our galaxy varies from place to place,
the same countrate limit corresponds to different
0.5-2 keV flux limits based on different galactic $N_{\rm H}$
values. The $N_{\rm H}$ value range from $(0.5-16)\times 10^{20}$ 
$[{\rm cm}^{-2}]$.  

\subsection{The RASS Selected-Area Survey -- North}

 This survey uses several high galactic latitude areas
of RASS (a total of $685 {\rm [deg^2]}$)  for 
optical identification of the sources 
down to about an order of magnitude fainter than the RBS.
The fields selected for the survey have 
the Galactic column ranging $N_{\rm H}=(2-11)\times 10^{20}$
$[{\rm cm}^{-2}]$. Details of the survey have been described 
in Zickgraf et al. (\cite{zick97}) and the catalog of source 
identifications has been published by Appenzeller et al. 
(\cite{app98}). We have further selected our sample
such that each field has a complete  {\it ROSAT}
hard-band (0.5-2 [keV]) countrate-limited 
sample with complete identifications 
($CR_{\rm hard} >$ 0.01-0.05 $[{\rm cts\;s^{-1}}]$).   

\subsection{The RIXOS Survey}

 The {\it ROSAT} International X-ray Optical Survey (RIXOS), Mason et al.
(\cite{rixos}) (see also Page et al. \cite{page96}) is a serendipitous
survey of $\approx 80$ PSPC fields covering  15 deg$^2$ of the sky. 
The flux limit of the deepest field is $S_{\rm x14}= 3.0$, while the
actual completeness limit varies from field to field.
The identification is $97\%$ complete, thus the effect of
the identification incompleteness is negligible considering statistical
errors.
   
\subsection{The North Ecliptic Pole (NEP) Survey}

 The data are from Bower et al. (\cite{nep}), which gave a catalog
of 20 sources in the 15$\farcm$5 radius region with 
$S_{\rm x14}\geq 1.0$. One object, RX J1802.1+6629 did not
have a redshift entry in Table 2 of Bower et al., but in the text,
they argued that the most probable interpretation of this object  
was a weak-lined QSO at $z\sim 1$. Thus we have assigned a redshift of
1.0 to this source. There is one unidentified source, making
the identificaton of the sample 95\% complete.  Thus we have set the 
effective survey area of the NEP survey as 95\% of the geometrical area. 

\subsection{The UK Deep Survey}
\label{sec:ukdsamp}

 Based on a 115 [ks] of {\it ROSAT} PSPC observation, McHardy et al. (1998) 
published a list of sources and identification of
X-ray sources down to $S_{\rm x14}=0.19$. A significant fraction 
of their identifications are ``NELGs'' (Narrow Emission-Line Galaxies)
and the fraction increases  towards fainter fluxes. 
{As mentioned in Sect. \ref{sec:intr}, a part of these NELGs
are likely to be misidentifications. The identifications
of other NELGs might be correct, but  
those would have been classified as AGNs with the criteria 
of Schmidt et al. (\cite{rds2}).}

 To include the results of the UKD survey in our sample, we would like
to include their NELGs in the latter category, but exclude those 
in the former category. We find that the redshift distribution of 
the AGN+NELG classes in the UKD survey is significantly different from 
the AGN+galaxy classes in the Lockman survey if we include all sources
down to $S_{\rm x14}\geq 0.19$. If we limit the sample to brighter sources 
($S_{\rm x14}\geq 0.5$), the redshift distributions are consistent with 
each other. Thus, in this work, we limit the samples from UKD and 
other deep PSPC surveys to $S_{\rm x14}\geq 0.5$, assuming that the 
misidentification problem would not affect the analysis significantly above
this limit.  

\subsection{The {\it ROSAT} Deep Survey -- Marano Field}

For the same reason as the UKD case, we have also used the same 
flux cutoff $S_{\rm x14}\geq 0.5$ for the survey in the 15$\arcmin$-
radius region on the  Marano field (Zamorani et al. \cite{marano}), based on 
a deep  PSPC exposure. Source fluxes of their catalog have been updated
since the version used by M99a.  The identifications are 100\% complete
for the 14 sources in $S_{\rm x14}\geq 1$ and 4 of the 27 sources remain 
unidentified or ambiguous (85\% complete) in 
$0.5\leq S_{\rm x14} <1$. As before, we have reduced the survey area 
by 15\% in this flux range to define the effective survey area used in
the SXLF calculations.
  
\subsection{The {\it ROSAT} Deep Survey -- Lockman Hole}
\label{sec:lhsamp}

 There are 200 ks of PSPC and 1 Msec of HRI observations
on this field (Hasinger et al. \cite{rds1}). The source list
and identifications for the brightest 50 sources
($S_{\rm x14}>0.5$) have been published (Schmidt et al. \cite{rds2}). 
In this work, we have included further unpublished identifications 
down to $S_{\rm x14}=0.17$. These include identifications and
redshifts based on spectra obtained in March 1998 with 
the Keck 10m telescope (Hasinger et al. \cite{has99}), which have 
also been included in M99a. Further four spectroscopic identifications
obtained with the Keck telescope in February 1999 have been added to
the sample since M99a. 

 The conversion between the HRI countrate and the 0.5-2 keV flux
has been determined from the mean values of overlapping
sources between the HRI and PSPC. The convsersion carries more
uncertainties based on spectra, because 
the HRI has practically no spectral resolution and has some
sensitivity down to 0.1 keV. With the HRI, the conversion factor varies by
$\pm 40\%$ for photon indices $\Gamma=2.0\pm 0.7$.   

The basic strategy of defining the combined PSPC-HRI sample
has been explained in Hasinger et al. (\cite{has99}). In this
paper, we have slightly modified the flux-limit and areas of the 
HRI sample in order to optimize our AGN sample in the presence of 
new identifications:  

\begin{itemize}
\item We use the deeper HRI-detected sample and HRI fluxes
 for the region 12.0 arcminutes from the HRI center (0.126 deg$^2$).
 At the faintest fluxes ($0.17\le S_{\rm x14} < 0.24$), we have
 further limited the area to 10.1 arcminutes from the HRI center
 (0.090 deg$^2$). This choice allows us to avoid the problem
 of incomplete source detection  due to source confusion 
 (see Fig. 2b of Hasinger et al. \cite{has99}).  A total of 48 AGNs 
 are present in this HRI sample.
    
\item Outside of the HRI region defined above, and within 18.4
 arcminutes from the PSPC center, we have used the PSPC detected
 sources and PSPC fluxes. This corresponds to 0.175 deg$^2$.
 For completeness, we have imposed a flux cutoff of 
 $S_{\rm x14}\ge 0.38$ for PSPC off-axis angles smaller 
 than 12$\arcmin$.5 and $S_{\rm x14}\ge 0.97$ for PSPC 
 offaxis angles between 12.5 and 18.4 arcminutes respectively.   

\item The sources in the HRI/PSPC combined sample have been 
 100\% identified for $S_{\rm x14}\ge 0.38$.
 Four of the 31 sources in $0.17\leq S_{\rm x14}< 0.38$ remain 
 spectroscopically unidentified.  Thus we have reduced the effective
 survey area by 13\% for $0.17\leq S_{\rm x14}<0.38$ to compensate 
 for the identification incompleteness.
 
\end{itemize}

\begin{figure}
\resizebox{\hsize}{!}{\includegraphics{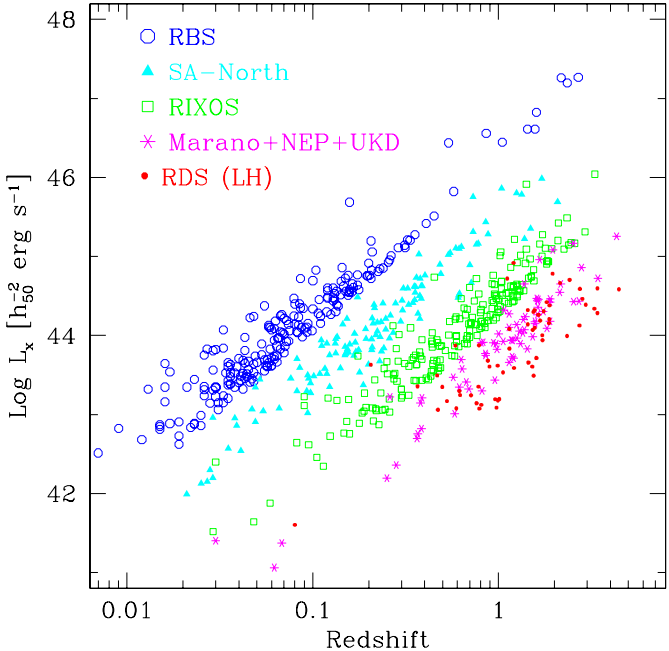}}
\caption[]{The AGNs in the combined sample are plotted
in the $z - Log L_{\rm x}$ for ($\Omega_{\rm}, \Omega_{\Lambda}$)
=(1.0,0.0). Different symbols correspond to different surveys 
as labeled.}
\label{fig:zl}
\end{figure}

\subsection{The combined sample}

 In our combined sample, there are 691 AGNs ranging from 
$4.2 \times 10^{-11}$ to $1.7\times 10^{-15} [{\rm erg\,s^{-1}\,cm^{-2}}]$.
The effective survey area for the combined sample is plotted as a function 
of the limiting flux in Fig. \ref{fig:area}, overlaid with
the value of $N(>S)^{-1}$, showing that the combined sample indeed
covers the above flux range continuously.   

 The redshift - luminosity scatter digrams of the
sample objects are shown in Fig. \ref{fig:zl} for 
the $(\Omega_{\rm 0},\Omega_{\rm \Lambda})=(1.0,0.0)$ universe. 
In any case, the luminosities have been calculated by:

\begin{equation}
L_{\rm x} = 4\pi d_{\rm L}(z)^2 S_{\rm 0.5-2 keV}
\label{eq:lx}
\end{equation}  

where $d_{\rm L}(z)$ is the luminosity distance as
a function of redshift, which depends on the
choice of cosmological parameters.
This corresponds to the no K-correction case. Explanation
on our K-correction policy is explained in Sect. \ref{sec:kcorr}.
Hereafter, the symbol $L_{\rm x}$ refers to the quantity defined 
in Eq. (\ref{eq:lx}) expressed in units of 
$h_{50}^{-2}{\rm [erg\,s^{-1}]}$, unless otherwise noted.

\section{The {\it ROSAT} AGN SXLF}

\subsection{K-Correction and AGN subclasses}
\label{sec:kcorr}

{In this section, we choose to present the SXLF in the
{\it observed} 0.5-2 keV band, i.e., in the $0.5(1+z)-2(1+z)$
keV range in the object's  rest frame. Thus no K-correction 
has been appplied for our expressions presented in this
section.} Also we choose to include
all emission-line AGNs (i.e., except BL-Lacs), 
including type 1's and type 2's. {The primary reason
for these choice is to separate the model-independent 
quantities, directly derived from {\it ROSAT} surveys,
from model-dependent assumptions.} Here we 
explain the philosophy behind these choices {in detail}. 

 There are a variety of AGN spectra in the X-ray
regime, but the information on exact content of AGNs in various
spectral classes is very limited.  Currently popular  
models explaining the origin of the 1-100 keV CXRB involve 
large contribution of self-absorbed AGNs (Madau et al.\cite{madau94}; 
Comastri et al. \cite{coma95}; Miyaji et al. \cite{m99b}; 
Gilli et al. \cite{gilli99}). Although they are selected
against in the {\it ROSAT} band, some of these absorbed AGNs come into
our sample. These absorbed AGNs certainly have different K-correction
properties than the unabsorbed ones. While these absorbed AGNs are mainly
associated with those optically classified as type 2 AGNs, the 
correspondence between the optical classification and the X-ray 
absorption is not straightforward. Especially, there are many
optically type-1 AGNs (with broad-permitted emission lines), which 
show apparent X-ray absorption of some kind. For example, a number 
of Broad Absorption Line (BAL) QSOs are known to have strongly
absorbed X-ray spectra 
(e.g. Mathur et al. \cite{phl5200}; Gallagher et al. \cite{gallag}). 
At the fainter/high-redshift end of our survey, there
may be some broad-line QSOs of this kind or some intermediate class.
{Broad-line AGNs with hard X-ray spectra have been found
in a number of hard surveys (Fiore et al. \cite{fiore}; Akiyama et
al. \cite{aki}).}   In Schartel et al. (\cite{schar97})'s study, all 
except two of the 29 AGNs from the 
Piccinotti et al.'s (\cite{picci}) catalog have been classified 
as type 1's, but about a half of them show X-ray absorption, 
some of which might be caused by warm absorbers. 
In view of these, using only optically-type 1 AGNs to exclude 
self-absorbed AGNs is not appropriate. Also optical classification
of type 1 and type 2 AGNs depend strongly on quality of optical 
spectra. Thus classification may be biased, e.g. as a function
of flux. However, the SXLF for the type 1 AGNs is of  
historical interest and shown in Appendix A. {As shown in
Appendix A., non type-1 AGNs are very small fraction of the 
total sample and excluding these does not change the main results
significantly.}  
   
 On the other hand, our sample of 691 AGNs with 
extremely high degree of completeness carries little uncertainties 
in the fluxes in the 0.5-2 keV band in the observer's frame, 
redshifts, and classification as AGNs. Thus,  we choose to show 
the SXLF expression in the observed 
0.5-2 keV band, or 0.5(1+z)-2(1+z) keV band at the source rest frame, 
in order to take full advantage of 
this excellent-quality sample without involving major sources of 
uncertainties. The expressions in the observed band may have less 
direct relevance for discussion on the actual AGN SXLF evolution.  
However they are more useful for discussing the contribution of 
AGNs to the Soft X-ray Background
(Sect. \ref{sec:cxrb}), interpretation of the fluctuation of the soft
CXRB, and evaluating the selection function for studying clustering 
properties of soft X-ray selected sample AGNs. 

 In practice, the expressions can also be considered a K-corrected SXLF
at the {\em zero-th approximation}, since  applying no K-correction 
is equivalent to a K-correction assuming $\Gamma=2$. 
This index has been historically
used in previous works (e.g. Maccacaro et al. \cite{emss1};
Jones et al. \cite{jones96}), thus our expression is useful for
comparisons with previous results. A $\Gamma=2$ power-law spectrum can be 
considered the best-bet single spectrum  characterizing the sample, 
because in the {\it ROSAT} sample, absorbed AGNs (including type 2 AGNs,
type 1 Seyferts with warm absorbers, BAL QSOs) are highly selected
against.   Nearby type 1 AGNs show an underlying power-law index of 
$\Gamma=2$ at $E\ga 1$[keV] (e.g. George et al. \cite{george}), 
which is the energy range corresponding to 0.5-2 keV for the
high redshifts where K-correction becomes important. The reflection 
component, which makes the spectrum apparently harder, becomes 
important only above 10 keV. This is 
outside of the {\it ROSAT} band even at $z\ga 4$.
The above argument is consistent with the fact that the average spectra of 
the faintest X-ray sources,
especially those indentified with broad-line AGNs, have 
$\Gamma \approx 2$ (Hasinger et al. \cite{has93}; 
Romero-Colmenero et al. \cite{rom}; Almaini et al. \cite{alm})
in the {\it ROSAT} band.
 Therefore, at the zero-th approximation, one can view our
expression as a K-corrected SXLF of AGNs, especially at high 
luminosities. The goodness of this approximation is highly model-dependent
and a discussion on further modeling beyond this zero-th approximation is  
given in Sect. \ref{sec:disc}. 
%
%

\begin{figure*}
\resizebox{\hsize}{!}{\includegraphics{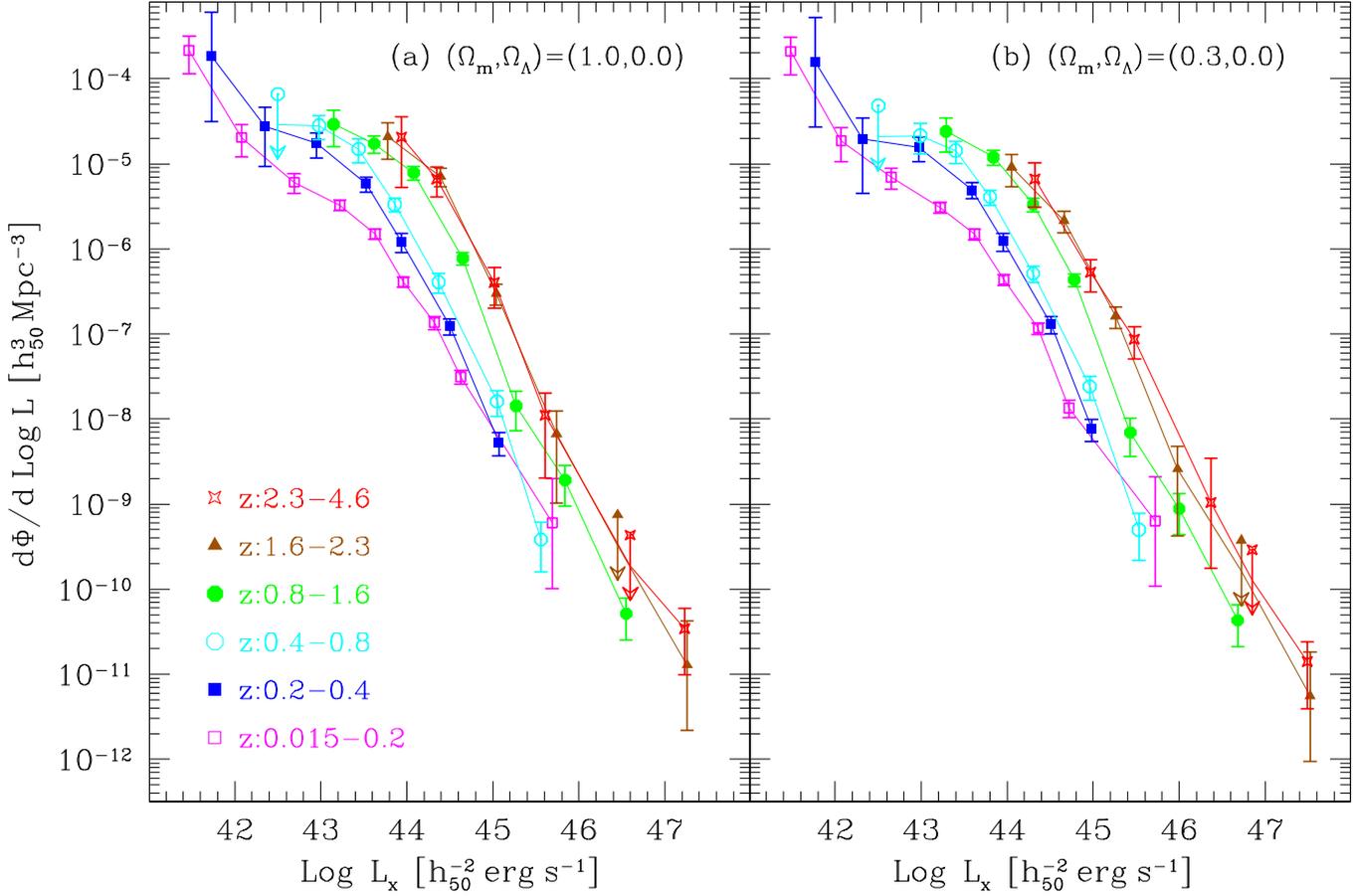}}
\caption[]{The $\Sigma V_{\rm a}^{-1}$ estimates of the SXLFs
are plotted with estimated 1$\sigma$ errors. Different
symbols correspond to different redshift bins as
indicated in the panel (a) and data points belonging
to the same redshift bin are connected. 
{The position
of the symbol attached to a downward arrow indicates
the 90\% upper limit (corresponding to 2.3 objects), where
there is no AGN detected in the bin.}}
\label{fig:xlf2}
\end{figure*}

\subsection{The binned SXLF of AGNs}
\label{sec:binxlf}

  The SXLF is the number density of soft X-ray-selected AGNs per 
unit comoving volume per ${\rm Log}\;L_{\rm x}$ as a function 
of $L_{\rm x}$ and $z$. We write the SXLF as:   
$$
\frac{{\rm d}\Phi} {{\rm d \;Log}\;L_{\rm x}}
({\rm Log}\; L_{\rm x},z).
$$

 Fig.\ref{fig:xlf2} shows the binned SXLF 
in different redshift shells estimated using the 
$\sum 1/V_{\rm a}$ estimator:
\begin{equation}
\frac{{\rm d}\;\Phi }{{\rm d\;Log}\;L_{\rm x}}
(\overline{{\rm\;Log}\;L_{{\rm x}j}},\overline{z_j})
\approx \frac {\sum_i\;V_{\rm a}^j(L_{{\rm x}i})^{-1}}
{(\Delta\;{\rm Log}\;L_{\rm x})_j},
\end{equation}

where the $L_{\rm x} - z$ bins are indexed by $j$ and
AGNs in the sample  falling into the $j$-th bin are
indexed by $i$, $V_{\rm a}^j(L_{{\rm x}})$ 
is the available comoving volume in the redshift range of 
the $j-th$ bin where an AGN with luminosity $L_{\rm x}$
would be in the sample. The luminosity function is estimated at 
($\overline{{\rm Log}\;L_{{\rm x}}}_j$,$\overline{z_j}$), 
where a bar represents the $V_{\rm a}^{-1}$ weighted average over 
the AGNs falling into the $j$-th bin. 
Also $(\Delta\;{\rm Log}\;L_{\rm x})_j$ is the size of the $j-$th bin
in ${\rm Log}\;L_{\rm x}$. 

Rough 1$\sigma$ errors have been estimated by:
\begin{equation}
\sigma{\left[\frac{{\rm d}\;\Phi }{{\rm d\;Log}\;L_{\rm x}}
(\overline{{\rm Log}\;L_{\rm x}}_j,\overline{z_j})\right]} 
\approx \frac{\sqrt{\sum_i\;V_{\rm a}(L_{{\rm x}i})^{-2}}}
             {(\Delta\;{\rm Log}\; L_{\rm x})_j}.
\label{eq:sigxlf}
\end{equation}
In case there is only one AGN in the bin,
{we have plotted
error bars which correspond to the exact Poisson errors
corresponding to the confidence range of Gaussian 1$\sigma$. 
In this way, we can also avoid
infinitely extending error bars in the logarithmic plot.}  

 Fig. \ref{fig:xlf2}(a)(b) shows the binned SXLF calculated 
for $(\Omega_{\rm m},\Omega_{\rm \Lambda})=(1.0,0.0)$ and $(0.3,0.0)$
respectively. In Fig. \ref{fig:xlf2}, we have also plotted 
some interesting upper-limits, in case there is no object in the bin. 
In the figure, we show upper limits corresponding to 2.3 objects 
(90\% upper-limit). See caption for details.   

 We note that the binned $\sum 1/V_{\rm a}$ estimate can cause 
a significant bias, especially because the size of the bins 
tend to be large. For example, at low luminosity bins with
corresponding fluxes close to the survey limit, the value of 
$V_{\rm a}$ can vary by a large
factor within one bin.  Also the choice of the point in $L_{\rm x}$ 
space representative of the bin, at which the SXLF values are plotted, may 
change the impression of the plot significantly. Thus the SXLF estimates 
based on the binned $\sum 1/V_{\rm a}$ can be used to obtain a rough 
overview of the behavior, but should not be used for statistical 
tests or a comparison with models. {Full numerical values of the
binned SXLF including $\sum 1/V_{\rm a}$ values, 
improved estimations by a method similar to
that discussed by Page \& Carrera \cite{page99}, and 
the numbers of AGNs in each bin will be presented in
paper II.} 

 A number of features can be seen in the SXLF. As found previously,
our SXLF at low $z$ is not consistent with a single
power-law, but turns over at around ${\rm Log}\;L_{\rm x}\sim 
43-44$. The SXLF drops rapidly with luminosity beyond the break.
We see a strong evolution of the SXLF up to the $0.8\leq z<1.6$
bin, but the SXLF does not seem to show significant evolution between 
the two highest redshift bins. Figs. \ref{fig:xlf2} (a)(b) show 
that these basic tendencies hold for the two extreme sets
of cosmological parameters.    

\subsection{Analytical expression -- statistical method}
\label{sec:stat}

 It is often convenient to express the SXLF and its evolution
in terms of a simple analytical formula, in particular, when 
using as basic starting point of further theoretical models.
   
 Here we explain the statistical methods of parameter 
estimations and evaluating the acceptance of the models. 
A minimum $\chi^2$ fitting to the binned $1/V_{\rm a}$ estimate
is not appropriate in this case because it can only be applied to 
binned datasets with Gaussian errors and at least 20-25 objects 
per bin are required to achieve this. In our case, such a bin is typically 
as large as  a factor of 10 in $L_{\rm x}$ and a factor of two in $z$, thus
the results would change depending where in the $(z\,L_{\rm x})$ bin
the comparison model is evaluated. 

 The Maximum-Likelihood method, where we exploit the 
full information from each object without binning, is a 
useful method for parameter estimations 
(e.g. Marshall et al.  \cite{mar83}),
while, unlike $\chi^2$, it does not give absolute goodness
of fit. The absolute goodness of fit can be evaluated 
using the one-dimensional and two-dimensional Kolgomorov-Smirnov tests 
(hereafter, 1D-KS and 2D-KS tests respectively; Press et al.\cite{numrec}; 
Fasano \& Franceschini \cite{ff_2dks})
to the best-fit models.  

As our maximum-likelihood estimator, we define  
\begin{equation}
{\cal L} = -2\;\sum_i \ln \left[\frac{N({\rm Log}\;L_{{\rm x}i},z_i)}
 	{\int \int N({\rm Log}\;L_{{\rm x}},z){\rm d\; Log}L_{\rm x}\;
	{\rm d}z}\right],
\label{eq:ml}
\end{equation} 
where $i$ goes through each AGN in the sample and 
$N({\rm Log}\;L_{{\rm x}},z)$ is the expected number 
density of AGNs in the sample per logarithmic luminosity
per redshift, calculated from a parameterized 
analytic model of the SXLF:

\begin{eqnarray}
       \nonumber \\
N({\rm Log}\;L_{{\rm x}},z) = &
	\nonumber \\
\frac{{\rm d}\;\Phi^{\rm model}}
	{{\rm d\; Log}\;L_{\rm x}}\;d_{\rm A}(z)^2\;(1+z)^3 & \;c\,
 \frac{{\rm d}\tau}{{\rm d}z}(z) \cdot A(L_{\rm x}/d_{\rm L}^2),
\end{eqnarray} 
where $d_{\rm A}(z)$ is the angular distance, $d\tau/dz(z)$ is the 
differential look back time per unit $z$ (e.g. Boldt \cite{boldt87})
and $A(S_{\rm x})$ is the survey area as a function of limiting 
X-ray flux (Fig. \ref{fig:area}). Minimizing ${\cal L}$ with respect
to model parameters gives the best-fit model. Since 
$\Delta {\cal L}$ from the best-fit point varies as $\Delta \chi^2$,
we determine the 90\% errors of the model parameters
corresponding to $\Delta {\cal L}=2.7$. The minimizations have been
made using the MINUIT Package from the CERN Program Library
(James \cite{minuit}).
 
 Since the likelihood function Eq. (\ref{eq:ml}) used normalized 
number density, the normalization of the model cannot be determined 
from minimizing ${\cal L}$, but must be determined independently. 
We have determined the model normalization (expressed by a parameter $A$ 
in the next subsections) such that the total number of expected 
objects (the denominator of the right-hand side of Eq. (\ref{eq:ml})) 
is equal to the number of AGNs in the sample ($N^{\rm obs}$).

Except for the global normalization $A$, we have made use of the MINUIT 
command MINOS (see James \cite{minuit}) to serach for errors.
The command searches for the 
parameter range corresponding to $\Delta {\cal L}\leq 2.7$, where all 
other free parameters have been re-fitted to minimize ${\cal L}$ during the 
search. The estimated 90\% confidence error for $A$ is taken to be
$1.7A\;(\sqrt{N^{\rm obs}})^{-1}$ and does not 
include the correlations of errors with other parameters.
 
 The 1D-KS tests have been applied to the sample distributions
on the $L_{\rm x}$ and $z$ space respectively. The 2D-KS test 
has been made to the function $N({\rm Log}\;L_{{\rm x}},z)$. We have 
shown the probability that the fitted model is correct based on the 
1D- and 2D-KS tests. For the 2D-KS test, calculated probability 
corresponding to the $D$ value from the analytical formula is accurate 
when there are $\ga 20$ objects and the probabilities $\la 0.2$. 
If we obtain a probability $\ga 0.2$, the exact
value does not have much meaning but implies that the model and data 
are not significantly different and we can consider the model 
acceptable. We have searched for models which have acceptance 
probabilities greater than 20\% in all of the KS tests. 
Strictly speaking, the analytical probability from
the KS-test $D$ values are only correct for models given {\em a priori}. 
If we use paramters fitted to the data, this would overestimate the 
confidence level. A full treatment should be made with large Monte-Carlo
simulations (Wisotzki \cite{wis}), where each simulated sample is
re-fitted and the D-value is calculated. However, making such large
simulations just to obtain formally-correct probability of 
goodness of fit is not worth the required computational task. Instead,
we choose to use the analytical probability and set rather 
strict acceptance criteria.

\subsection{Analytical expression -- overall AGN SXLF}

 Using the method described above, we have searched for 
an analytical expression of the overall SXLF.
The overall fit has been made for the redshift range
$0.015\leq z<5$. Also for the fits, we have limited the 
luminosity range to ${\rm Log}\;L_{\rm x}\ga 41.7$.
 
 As described in Sect. \ref{sec:data}, the lower redshift cutoff is 
imposed to avoid effects of local large scale structures, which 
may cause a deviation from the mean density of the present epoch 
and thus can cause significant bias to the low luminosity behavior 
of the SXLF. 
 At the lowest luminosities (${\rm Log}\;L_{\rm x} \le 41.7$), 
there is a significant excess of the SXLF from the extrapolation from 
higher luminosities. This excess connects well with the nearby 
galaxy SXLF by Schmidt et al. (\cite{sbv})
(see also e.g. Hasinger et al. \cite{has99}) and may well contain contamination
from star formation activity (see also Lehmann et al. \cite{leh_tru}).
For finding an analytical overall
expression, we have not included the AGNs belonging to this
regime. 
     
 As an analytical expression of the present-day ($z=0$) SXLF, we
use the smoothly-connected two power-law form:

\begin{equation}
\frac{{\rm d}\;\Phi\,(L_{\rm x},z=0) }{{\rm d\;Log}\;L_{\rm x}}
  = {A}\;\left[(L_{\rm x}/{L_*})
      ^{{\gamma_1}}
         +(L_{\rm x}/{L_*})^{{\gamma_2}}
  	\right]^{-1}
\end{equation}

 As a description of evolution laws, the following models 
have been considered:

\subsubsection{Pure-luminosity and pure-density evolutions}
\label{sec:ple_pde}

As some previous works (e.g. Della Ceca et al.
\cite{emss2}; Boyle et al. \cite{boyle94}; Jones et al. \cite{jones96};
Page et al. \cite{page96}), we have first tried to fit the SXLF
with a pure-luminosity evolution (PLE) model.

\begin{equation}
\frac{{\rm d}\;\Phi\,(L_{\rm x},z)}{{\rm d\;Log}\;L_{\rm x}}
  = \frac{{\rm d}\;\Phi\,(L_{\rm x}/e_{\rm}(z),0)}
    {{\rm d\;Log}\;L_{\rm x}}
\label{eq:ple}
\end{equation}

\begin{table}
\caption[]{Best-fit PLE and PDE Parameters} 
\begin{center}
\begin{tabular}{ll}
\hline\hline
Model    & Parameters$^{\rm a}$/KS probabilities\\
           $(\Omega_m,\Omega_\Lambda)$ \\
\hline
\\
PLE & $A=(4.0 \pm .3)\times 10^{-6}$;\,$L_*=0.33\pm.10$ \\
(1.0,0.0)& $\gamma_1=0.60\pm .16$;\,$\gamma_2=2.34\pm .12$;\,$p1=3.0\pm .2$\\
     & $z_{\rm c}=1.42\pm .17$;\,$p2=0.3^{+0.5}_{-1.0}$\\ 
     & \underline{$P_{\rm KS}=.002,3\; 10^{-5}, 1\; 10^{-5}$ (for $L$,$z$,2D)}\\
\\
PLE & $A=(3.1 \pm .2)\times 10^{-6}$;\,$L_*=0.38\pm.12$ \\
(0.3,0.0) & $\gamma_1=0.57\pm .16$;\,$\gamma_2=2.35\pm .12$;\,$p1=2.9\pm .2$\\
     & $z_{\rm c}=1.54\pm .25$;\,$p2=0.3\pm .7$\\
     & \underline{$P_{\rm KS}=.08,.001,2\; 10^{-4}$ 
     (for $L$,$z$,2D)}\\
\\
PDE & $A=(6.0 \pm .4)\times 10^{-7}$;\,$L_*=1.08\pm .4$ \\
(1.0,0.0)& $\gamma_1=0.74\pm .13$;\,$\gamma_2=2.28\pm .11$;\,$p1=4.6\pm .3$\\
     & $z_{\rm c}=1.60\pm .25$;\,$p2=0.6\pm 1.1$\\ 
     & \underline{$P_{\rm KS}=0.9,0.9,0.16$ (for $L$,$z$,2D)}\\
\\
PDE & $A=(5.4 \pm .3)\times 10^{-7}$;\,$L_*=1.13\pm 0.4$ \\
(0.3,0.0)& $\gamma_1=0.76\pm .13$;\,$\gamma_2=2.22\pm .10$;\,$p1=4.6\pm .3$\\
     & $z_{\rm c}=1.62\pm .26$;\,$p2=1.3\pm 1.1$\\ 
     & \underline{$P_{\rm KS}= {0.8,0.7,0.1}$ (for $L$,$z$,2D)}\\
\\
\hline
\end{tabular}
\end{center}

$^{\rm a}$Units -- A: [$h_{50}^3\;{\rm Mpc^{-3}}$],\,\,
 $L_*$: [$10^{44}\;h_{50}^{-2}{\rm erg\;s^{-1}}$],\,\,
$I_{\rm x12}$: $[10^{-12}{\rm erg\,s^{-1}\,cm^{-2}\,deg^{-2}}]$ 
in 0.5-2 keV. Parameter errors correspond to the 90\% confidence
level (see Sect. \ref{sec:stat}).
\label{tab:ple_pde} 
\end{table}

For the evolution factor, we have used a power-law form:

\begin{equation}
e(z) = \left\{ 
	\begin{array}{ll}
	(1+z)^{p1} & (z \leq z_{\rm c}) \\ 
        e(z_{\rm c})[(1+z)/(1+z_{\rm c})]^{p2} & (z > z_{\rm c})\\
	\end{array}
       \right. .
\label{eq:ez}
\end{equation}

The best-fit values are listed in the upper
part of Table \ref{tab:ple_pde} along with 1D-KS and 2D-KS 
probabilities using the analytical formula. 
In Table \ref{tab:ple_pde} and later tables, 
the three values of $P_{\rm KS}$ represent the probabilities 
that the model is acceptable for the 1D-KS test in the $L_{\rm x}$ 
distribution, 1D-KS test in the $z$ distribution, and 2D-KS test in the
($L_{\rm x}$,$z$) distribution respectively. Note that there are
cases which are accepted by 1D-KS tests in both distributions but
fail in the 2D-KS test. The results of the fit show that
the PLE model is certainly rejected with a 2D-KS probability of
$P_{\rm 2DKS}= 5\times 10^{-5}$ and $1\times 10^{-2}$ for
the $\Omega_{\rm m}$=1 and 0.3 ($\Omega_\Lambda=0$) cosmologies 
respectively.

 As an alternative, we have also tried the Pure-Density 
Evolution model (PDE), which seemed to fit well in our 
preliminary analysis for the $\Omega_{\rm m}$=1 ($\Omega_\Lambda=0$)
universe (Hasinger \cite{has_xs}).

\begin{equation}
\frac{{\rm d}\;\Phi\,(L_{\rm x},z)}{{\rm d\;Log}\;L_{\rm x}}
  = \frac{{\rm d}\;\Phi\,(L_{\rm x}/,0)}
    {{\rm d\;Log}\;L_{\rm x}}\cdot e(z)
\label{eq:pde}
\end{equation}

where $e(z)$ has the same form as Eq. (\ref{eq:ez}). The 2D-KS 
probabilities are $P_{\rm 2DKS}= 0.16$ and
{0.1} for the $\Omega_{\rm m}$=1 and 0.3 ($\Omega_\Lambda=0$) 
respectively. Thus the acceptance of the 
overall fit is marginal, especially for  $\Omega_{\rm m}$=1. 
However the PDE model has a serious problem of overproducing the 
soft X-ray background (Sect. \ref{sec:cxrb}). 
For a further check, we have made
separate fits to high luminosity (${\rm Log}\;L_{\rm x}>44.0$)
and low luminosity (${\rm Log}\;L_{\rm x}>44.0$) samples 
to compare the evolution index $p1$ in
$0.015<z<1.6$ for $\Omega_{\rm m}=1.0$. We have obtained 
$p1=5.3\pm 0.5$ and $4.1\pm 0.5$ (90\% errors) for the high 
and low luminosity samples respectively.
Thus the density evolution rate is somewhat slower at
low luminosities. Of course at the low luminosity regime, the fit was 
weighted towards nearby objects. If the evolution does not exactly follow 
the power-law form ($\propto [1+z]^{p1}$), spurious difference in 
evolution rate can arise. {Visual inspection of
Fig. \ref{fig:xlf2} might suggest that at $z<0.4$, the evolution rate 
seems larger at low luminosities, as opposed to the results shown
above for $z<1.6$. However, performing the same experiment for 
the $z<0.4$ AGNs showed $p1=5.7\pm 1.8$ and $5.8\pm 1.2$ for the 
high and low luminosity samples respectively, indicating no difference
within relatively large errors. For the $0.4\geq z >1.6$ sample,
the results are  $p1=6.2\pm 0.8$ and $3.0\pm 1.0$, again, for the 
high and low luminosity samples respectively.}
This difference  and the soft CXRB 
overproduction problem lead us to explore a more sophisticated form
of the overall SXLF expression as described in the next section.
    
\subsubsection{Luminosity-dependent density evolution}
\label{sec:ldde1}

\begin{table}
\caption[]{Best-Fit LDDE1 Parameters} 
\begin{center}
\begin{tabular}{ll}
\hline\hline
Model    & Parameters$^{\rm a}$/KS probabilities \\
 $(\Omega_m,\Omega_\Lambda)$ \\
\hline
\\
LDDE1 & $A=(1.01 \pm .06)\times 10^{-6}$;\,$L_*=0.75^{+.41}_{-.26}$ \\
(1.0,0.0) & $\gamma_1=0.75\pm .15$;\,$\gamma_2=2.25\pm .10$;\,$p1=5.1\pm .3$\\
     & $z_{\rm c}=1.57\pm .15$;\,$p2=0.0$ (fixed) \\
     & $\alpha = 1.7\pm .8$;\, ${\rm Log} L_{\rm a}=44.1$ (fixed) \\
     & \underline{$P_{\rm KS}=0.6, 0.4, 0.5$ (for $L$,$z$,2D)};\,\\
\\ 	
LDDE1 & $A=(1.56\pm .10)\times 10^{-6}$;\,$L_*=0.56^{+.33}_{-.18}$ \\
(0.3,0.0) & $\gamma_1=0.68\pm .18$;\,$\gamma_2=2.19\pm .08$;\,$p1=5.3\pm .4$\\
     & $z_{\rm c}=1.59\pm .14$;\, $p2=0.0$ (fixed) \\
     & $\alpha = 2.3\pm .7$;\, ${\rm Log} L_{\rm a}=44.3$ (fixed) \\ 
     & \underline{$P_{\rm KS}=0.5,0.3,0.3$ (for $L$,$z$,2D)}\\

\\
LDDE1 & $A=(1.61\pm .10)\times 10^{-6}$;\,$L_*=0.56^{+.32}_{-.18}$ \\
(0.3,0.7) & $\gamma_1=0.66\pm .18$;\,$\gamma_2=2.19\pm .08$;\,$p1=5.3\pm .4$\\
     & $z_{\rm c}=1.58\pm .14$;\, $p2=0.0$ (fixed) \\
     & $\alpha = 2.6\pm .7$;\, ${\rm Log} L_{\rm a}=44.4$ (fixed) \\ 
     & \underline{$P_{\rm KS}=0.4,0.4,0.3$ (for $L$,$z$,2D)}\\
\\
\hline
\end{tabular}
\end{center}
$^{\rm a}$Units -- A: [$h_{50}^3\;{\rm Mpc^{-3}}$],\,\,
 $L_*$: [$10^{44}\;h_{50}^{-2}{\rm erg\;s^{-1}}$],\,\,
Parameter errors correspond to the 90\%
confidence level (see Sect. \ref{sec:stat}).
\label{tab:ldde1_fit} 
\end{table}

   We have tried a more complicated 
description by modifying the PDE model such that the evolution 
rate depends on luminosity  (the Luminosity-Dependent Density 
Evolution model). In particular, as shown above, it seems that 
lower evolution rate at low luminosities than the PDE case 
would fit the data well. This tendency is also seen in the optical
luminosity function of QSOs (Schmidt \& Green \cite{schm_gre};
Wisotzki \cite{wis}). The particular form we have first tried (the LDDE1
model) replaces $e(z)$ in Eq. (\ref{eq:pde}) by $e(z,L_{\rm x})$, where 

\begin{eqnarray}
  \lefteqn{e(z,L_{\rm x}) = } \nonumber \\
      & \left\{ 
	\begin{array}{ll}
	  (1+z)^{\max(0,{p1}-{\alpha}\,{\rm Log}\; 
	  {[L_{\rm a}/L_{\rm x}]})} 
	     & (z \leq z_{\rm c}; L_{\rm x}<L_{\rm a}) \\ 
	  (1+z)^{p1}
	     & (z \leq {z_{\rm c}}; L_{\rm x}\ge L_{\rm a})\\ 
        e({z_{\rm c}},L_{\rm x})
         \left[(1+z)/(1+{z_{\rm c}}) \right]^{p2} 
	     & (z>{z_{\rm c}}) \\
	\end{array}
       \right.
\label{eq:lddez}
\end{eqnarray}

 In Eq. (\ref{eq:lddez}), The parameter $\alpha$ represents the degree of 
luminosity dependence on the density evolution rate for 
$L_{\rm x}<L_{\rm a}$. The PDE case is $\alpha=0$ and a greater 
value indicates lower density evolution rates at low luminosities. 

 The best-fit LDDE1 parameters, the results of the KS tests, and the
integrated 0.5-2 keV intensity are shown in Table \ref{tab:ldde1_fit}.
Table \ref{tab:ldde1_fit} shows that considering the luminosity 
dependence to the density evolution law has significantly improved
the fit. The 2D-KS probabilities (analytical) are more than 30\% for 
all sets of cosmological parameters. 

We have considerd another form of the LDDE model (designated
as LDDE2), which was made to produce 90\% of the estimated
0.5-2 keV extragalactic background. The details of the construction of
the LDDE2 is discussed in Sect. \ref{sec:cxrb}, where  
the contribution to the Soft Cosmic X-ray Background is discussed. 
In figures in the following discussions, 
the LDDE2 model is also plotted.  

\begin{figure}
\resizebox{\hsize}{!}{\includegraphics{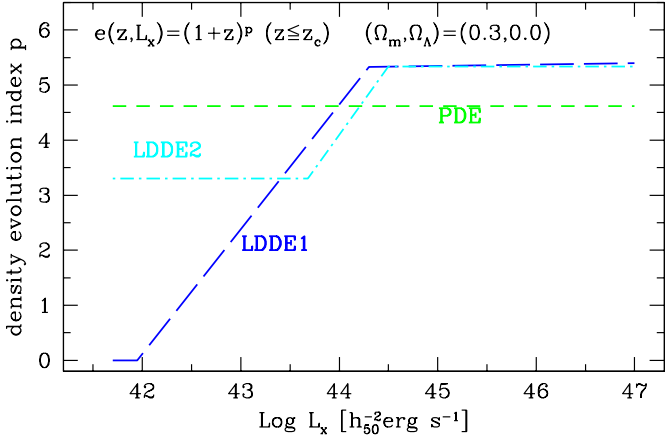}}
\caption[]{The behavior of the evolution indices at $z\la z_{\rm c}$
are shown as a function of luminosity for various density evolution 
models: PDE (short-dashed, Sect. \ref{sec:ple_pde}), LDDE1 (long-dashed 
\ref{sec:ldde1}), and LDDE2 (dot-dashed, Sect. \ref{sec:cxrb}).
The lines for the  $(\Omega_{\rm m},\Omega_{\rm \Lambda})=$ $(0.3,0)$
case are shown.}
\label{fig:pev}
\end{figure}

\begin{figure}
\resizebox{\hsize}{!}{\includegraphics{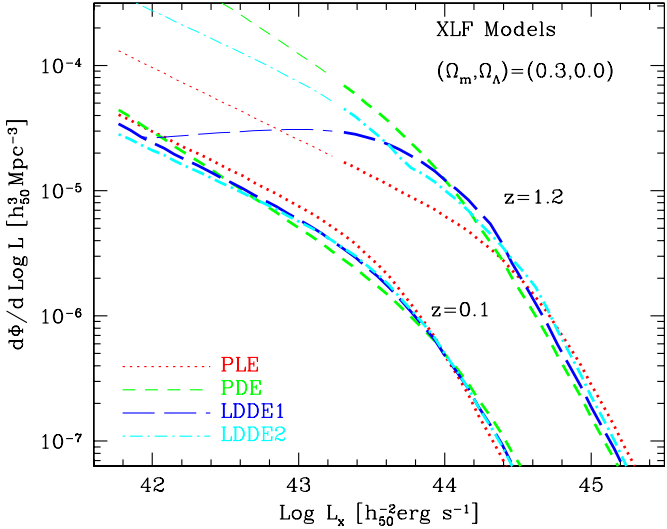}}
\caption[]{The behavior of the model SXLFs at z=0.1 and 1.2
are shown respectively for the PLE (dotted), 
PDE (short-dashed), LDDE1 (long-dashed), and LDDE2 (dot-dashed) models.
For the z=1.2 curves, thick-line parts show the portion covered
by the sample ($S_{\rm x14}\ga 0.2$) and the thin-line parts are
extrapolations to fainter fluxes.  The lines are for  
$(\Omega_{\rm m},\Omega_\Lambda)=$ $(0.3,0)$.}
\label{fig:xlfm}
\end{figure}

 For an illustration, in Fig. \ref{fig:pev} we show the behavior of the density
evolution index for $z\leq z_{\rm c}$ as a function of luminosity 
for our PDE, LDDE1 and LDDE2 models. Fig \ref{fig:xlfm} shows the 
behavior of the model SXLFs at z=0.1 and 1.2. In this figure, only
the part drawn in thick lines is constrained by data and thin
lines are model extrapolations. These figures are only meant for 
illustrative purposes and thus are only shown for the 
$(\Omega_m,\Omega_\Lambda)$=$(0.3,0)$ cosmology, where
differences among models are more pronounced.

\subsection{Comparison of the data and the models}
\label{sec:dmcom}

\begin{figure}
\resizebox{\hsize}{!}{\includegraphics{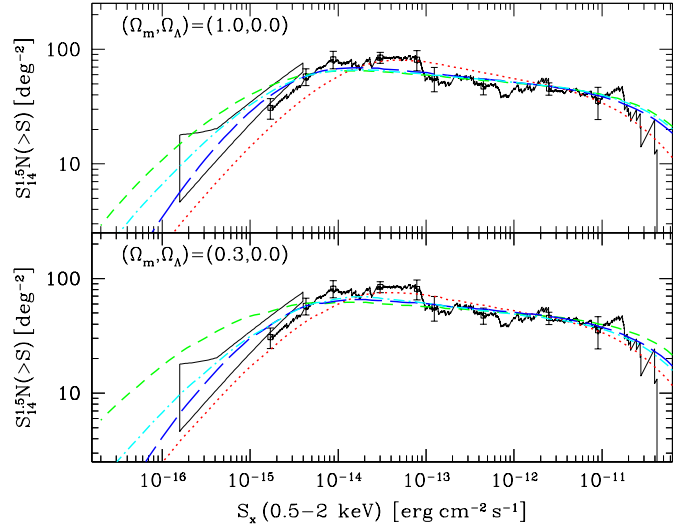}}
\caption[]{The $S^{1.5}N(>S)$ (a horizontal line corresponds to
the Euclidean slope) curve for our sample AGNs is plotted with 90\% 
errors at several locations and are compared with the best-fit PLE (dotted), 
PDE (shot-dashed), LDDE1 (long-dashed), LDDE2 (dot-dashed) models 
for the $(\Omega_{\rm m},\Omega_{\rm \Lambda})=$ $(1,0)$ (upper panel)
and $(0.3,0)$ (lower panel). The thin-solid fish is from the fluctuation 
analysis of the Lockman Hole HRI data (including non-AGNs) by
Hasinger et al. (\cite{has93})}
\label{fig:rlnls2}
\end{figure}

\begin{figure}
\resizebox{\hsize}{!}{\includegraphics{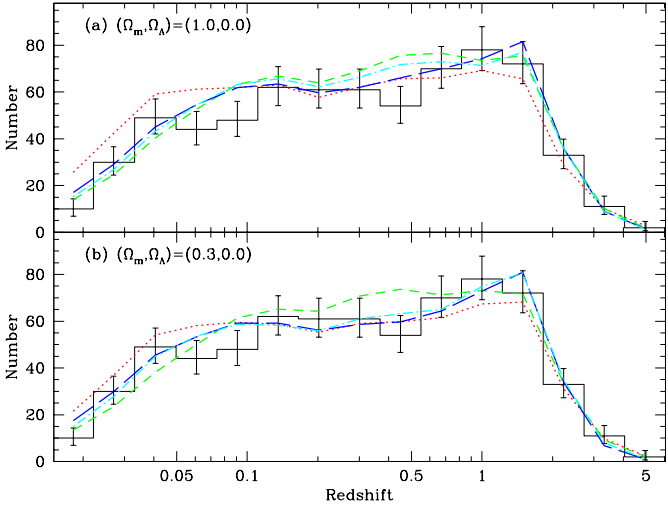}}
\caption[]{The redshift distribution of the AGN sample, histogrammed in 
equal interval in $\log\,z$, is compared with predictions
from the best-fit PLE (dotted), PDE (shot-dashed), LDDE1
(long-dashed), and LDDE2 (dot-dashed) models for two sets of 
cosmological parameters as labeled.
The assymmetric error bars correspond to approximate 1$\sigma$ Poisson
errors calculated using Eqs. (7) and (11) of 
Gehrels (\cite{gehrels}) with $S=1$.  }

\label{fig:zdist}
\end{figure}

 For a demonstration of the comparison between the
analytical expressions and the data, we have  plotted the
$S^{1.5}N(>S)$ curve (the Log $N$ -- Log $S$ curve plotted
in such a way that the Euclidean slope becomes horizontal)
for AGNs in our sample with expectations from our models 
(Fig. \ref{fig:rlnls2}). Also the redshift distribution of
the sample has been compared with the models in 
Fig. \ref{fig:zdist}. These two comparisons already show
intersting features. As expected, the PLE underpredicts
and PDE overpredicts the number counts of lowest flux sources.
In the redshift distribution, the PLE overpredicts the
number of $z\la 0.08$ sources while it slightly underpredicts
the $z\approx 1$ sources. Although the deviation in
each redshift bin seems small, the deviations in the 
neighboring bins are consistent and these systematic deviations
can be sensitively picked up by the KS test in the $z$ distribution 
(see small values of the $P_{\rm KS}$ in $z$ for the PLE model in 
Table \ref{tab:ple_pde}).

\begin{figure}
\resizebox{\hsize}{!}{\includegraphics{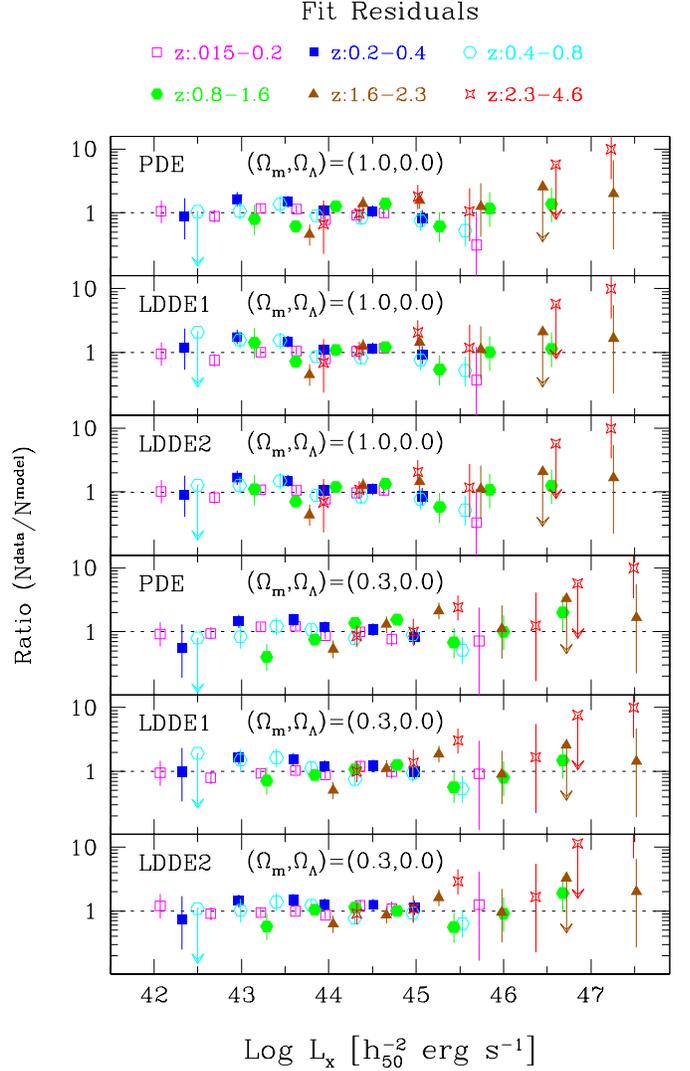}}
\caption[]{The full residuals of the fit are shown for the PDE, LDDE1 and
LDDE2 models in two sets of cosmological parameters as labeled in
each panel. The redidual in each bin has been calculated from actual number 
of sample AGNs falling into the bin and the model predicted number.  
Different symbols correspond to different redshidt bins 
as indicated above the top panel, which are identical to those
used in Fig.\ref{fig:xlf2}. One sigma errors have been plotted using
approximations to the Poisson errors given in Gehrels (\cite{gehrels}).
The upper limit corresponds 2.3 objects (90\% upper-limit).}

\label{fig:resid2}
\end{figure}

\begin{figure}
\resizebox{\hsize}{!}{\includegraphics{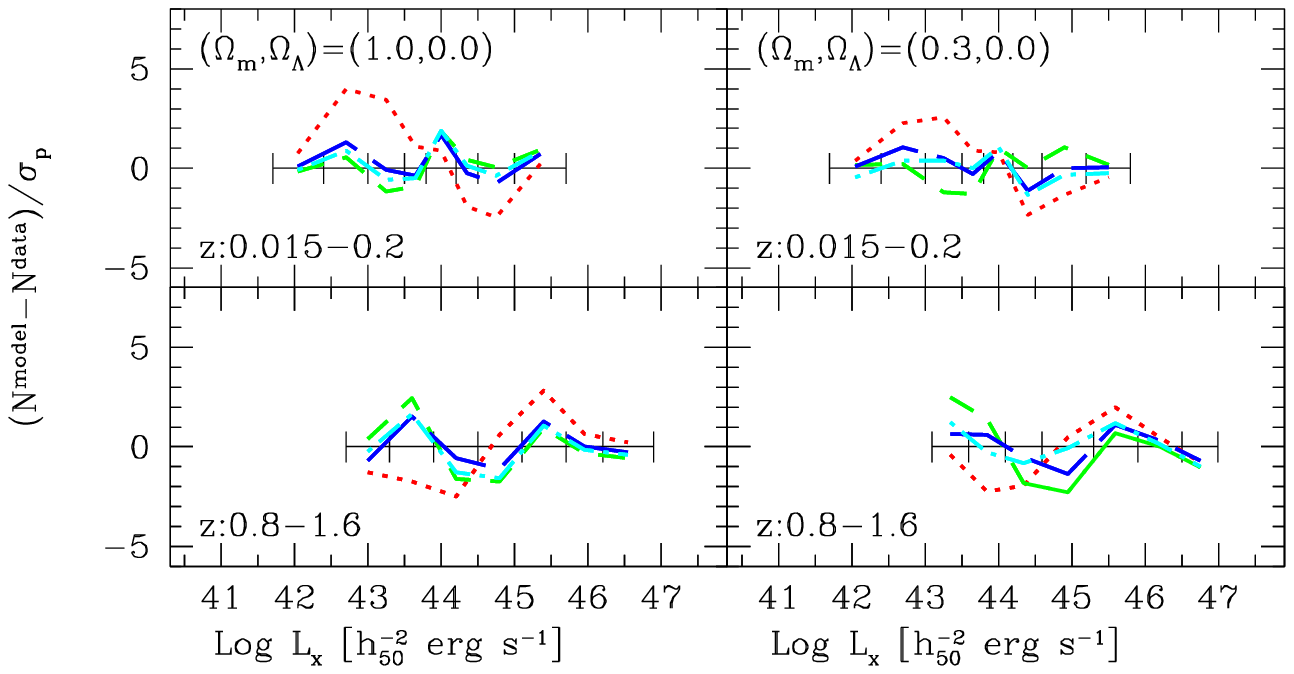}}
\caption[]{Residuals in the -$\chi$ space (see text) are shown
for two resdhift bins, i.e., 0.015$\leq z <0.2$ and 0.8$\leq z <1.6$, 
where differences among different models are apparent. Different 
line styles correspond to different models. See
caption for Fig. \ref{fig:xlfm} for the line styles.
The luminosity bins are shown as horizontal bars bordered by ticks.}

\label{fig:nsig2}
\end{figure}

 The plots in Figs. \ref{fig:rlnls2} and \ref{fig:zdist} are 
comparisons of distributions in one-dimensional 
projections of a two-dimensional distribution.
Only with these projected plots, one can easily overlook important
residuals localized at certain locations. Thus we 
also would like to show the comparison
in the full two-dimensional space. In literature, models are often 
overplotted to the binned SXLF plot calculated by the $V_{\rm a}^{-1}$
estimate like Fig. \ref{fig:xlf2}. However, given unavoidable
biases associated with the binned  $V_{\rm a}^{-1}$ estimates
(see Sect. \ref{sec:binxlf}), such a plot can cause one to pick up 
spurious residuals. Thus we have plotted residuals in the 
following unbiased manner. For each model, we have calculated the 
expected number
of objects falling into each bin ($N^{\rm model}$) and compared with the 
actual number of AGNs observed in the bin ($N^{\rm data}$). The full residuals
in term of the ratio $N^{\rm data}/N^{\rm model}$ are
plotted in Fig. \ref{fig:resid2} for
the {PDE}, LDDE1 and LDDE2 models for two sets of cosmological parameters
as labeled. The error bars correspond to 1$\sigma$ Poisson errors
($\sigma_{\rm p}$) estimated using Eqs. (7) and (11) of Gehrels 
(\cite{gehrels}) with $S=1$. Points belonging to different redshift 
bins are plotted using different symbols as labeled (identical to those in
Fig. \ref{fig:xlf2}). These residual plots show which part of the
$z-L_{\rm x}$ space the given models are most representative of,
which part is less constrained because of the
poor statistics, and where there are systematic residuals. It seems that the
models underpredict the number of AGNs in the highest luminosity bin
at $2.3\leq z< 4.6$ by a factor of 10, but statistical significance
of the excess is still poor (2 objects against the models predictions 
of about 0.2). These AGNs do not constrain the fit strongly and 
excluding them did not change the results significantly. 
Also there is a scatter up to a factor of 2 from the model in 
$45\la {\rm Log}\,L_{\rm x}\la 46$, but no points
are more than 2$\sigma$ away from either of the LDDE1 and LDDE2 
models in both cosmologies. 

The only data point which is more than 2$\sigma$ away from LDDE1 
or LDDE2 model is the lowest
luminosity bin at $1.6\leq z< 2.3$ (filled triangle), i.e., 
$43.6\la {\rm Log}\,L_{\rm x}\la 44.2$ for 
($\Omega_{\rm m}, \Omega_\Lambda$) = (1.0,0.0) or
$43.8\la {\rm Log}\,L_{\rm x}\la 44.5$ for 
($\Omega_{\rm m}, \Omega_\Lambda$) = (0.3,0.0).
Both LDDE1 and LDDE2 models overpredict the number of AGNs 
by a factor of $\approx 2$ in both cosmologies, which are 
$2.2-3.8\sigma$ away. However, this location corresponds to
the faintest end of the deep surveys with a certain amount of
incompleteness in the identifications. Our incompleteness
correction  method (Sect. \ref{sec:data}) is valid
only if the unidentified source are {\em random} selections of the
X-ray sources in the similar flux range.
However, these sources have remained unidentified because
of the difficulty of obtaining good optical spectra and not by
a random cause. Thus it is possible that the incompleteness
preferentially affects a certain redshift range. Actually the
deficiencies were much larger in the previous version 
(see Fig. 8 of Hasinger et al. \cite{has99}). The discrepancies
decreased after the February 1999 Keck observations of the faintest 
Lockman Hole sources with rather long exposures, where three of 
the four newly identified source turned out to be concentrated 
in this regime. Thus it is quite possible that the remaining
{four} unidentified sources are also concentrated in this regime.
{In that case, the LDDE models can also fit to this bin within
2$\sigma$.}
{Actually the newly identified and unidentified sources 
typically have very red $R-K^\prime$ colors (Hasinger et al. \cite{has99};
Lehmann et al. \cite{rds3}), which
probably belong to a similar class to those found by 
Newsam et al. (\cite{newsam}). If the red $R-K^\prime$ color
comes from the stellar population of underlying galaxy, they
are likely to be in a concentrated redshift regime. On the other hand,
if it represents obscured AGN component, they can be in a variety
of redshift range.} 
At this moment, it is not clear whether the 
deficiencies in this location is due to incompleteness or indicate 
an actual behavior of the SXLF.

 {Based on the results of the 1-D and 2-D KS tests, we have
rejected the PLE model. We favor the LDDE1 and LDDE2 models
over the PDE model based on the KS tests and as well as the CXRB 
constraints (see below).}
It may be interesting to show the exact location where the largest
discrepancies are for these models, as compared to the LDDE models.
This can be most clearly shown by plotting residuals in the
$-\chi = (N^{\rm model}-N^{\rm data})/\sigma_{\rm p}$ space. 
We have shown the $-\chi$ residuals for redshift bins where 
there are notable differences among these models,
i.e., $0.015\leq z < 0.2$ and $0.8\leq z < 1.6$. 
These are shown in Fig. \ref{fig:nsig2}. For both cosmologies,
the PLE model systematically shifts from overprediction to
underprediction with increasing luminosity at the lowest 
redshift bin. At the higher resdhift bin, the opposite shift
can be seen. The curve converges closer to zero
at both high and low luminosity ends just because there are only small
numbers of objects in these bins causing poor statistics.
More apparently in the $(\Omega_{\rm m},\Omega_\Lambda)=(0.3,0)$
universe, the PDE model also shows a significant scatter around zero.

{The data in the lower luminosity part 
$42 \la {\rm Log}\,L_{\rm x}\la 43.5$ in the lowest redshift 
bin ($0.015\leq z < 0.2$) are crucial in rejecting the PLE model, 
as seen in Figs. \ref{fig:xlf2} and \ref{fig:nsig2}. This regime,
consisting of $\sim 90$ AGNs, has low SXLF values compared with 
the PLE extrapolation from the higher redshift data. 
Actually we cannot discriminate between the PLE and LDDE models 
for the sample of AGNs with $z<0.2$ excluded. For the $z\ge 0.2$
sample, we could find good fits (with all of the KS
probabilities in $L_{\rm x}$, $z$, and 2D
exceeding 0.2) in any of the PLE and LDDE models. The acceptance
of the PDE model was marginal ($P_{\rm 2DKS}\sim 0.1$). 
The $z<0.2$ regime is mainly contributed by AGNs in 
the RASS-based RBS and SA-N surveys, 
whose flux-area space have not been explored previously.      
Since these samples are completely identified
(see Sect. \ref{sec:data}) and we have included all 
emission-line AGNs, the relatively low value in this 
regime is not because of the incompleteness or sampling 
effects. The only source of possible systematic errors which
could affect the analysis would be in the flux measurements, because
of the differences in details of the source detection methods among
different samples. Some systematic shift of flux measurements
might have occured between measurements in, e.g., 
the pointed and RASS data (for which there is no evidence). 
Thus we have made a sensitivity check by 
shifting the fluxes of all RBS and SA-N AGNs by $+20\%$ and
$-20\%$. The flux-area relation (Fig. \ref{fig:area}) has been
modified accordingly. In either case in either value of 
$\Omega_{\rm m}$, the basic results did
not change and especially the PLE model has been
rejected with a large significance 
(with $P_{\rm 2DKS}$ ranging  $10^{-3}-10^{-6}$).}

\section{Contribution to the Soft X-ray Background}
\label{sec:cxrb}

 In this section, we discuss the contribution of AGNs
to the soft X-ray background using the various models
of the SXLF. As the absolute intensity level of the extragalactic
0.5-2 keV CXRB intensity, we use the results of an {\it ASCA-ROSAT}
simultaneous analysis on the {\it ASCA} LSS field 
(Miyaji et al. in preparation), which covers a much larger 
field than Miyaji et al. (\cite{cxbsp}) and thus is subject
to less uncertainties due to source fluctuations. 
There still are uncertainties in separation of the Galactic hard thermal 
and extragalactic components. 
Especially, it is still not clear whether the extragalactic component 
has also a soft excess at $E\la 1$ [keV] over the extrapolation 
from higher energies or whether the observed excess is dominated by the
Galactic hard thermal component.
Some authors prefer a model where the extragalactic
component also contributes to the $E\la 1$ [keV] excess
because fit with a single power-law plus a thermal plasma would 
require an unusally low metal abundance of the thermal 
component for a Galactic plasma 
(Gendreau et al. \cite{gend}) and/or because many AGNs show soft 
excesses (e.g. Parmar et al. \cite{parmar}).
On the other hand, a self-consistent population synthesis
model, including the AGN soft-excess below 1.3 keV, still predicts 
that the low-energy excess is not prominent in the 0.5-2 keV range
(Miyaji et al. \cite{m99b}), mainly because the break energy
shifts to the observed photon enrgy of $E\sim 0.4$ [keV] for AGNs 
at $z\sim 2$, where the largest contribution to the CXRB is expected.
The 0.25 keV extragalactic component measured using a shadowing
of a few nearby galaxies (Warwick \& Roberts \cite{shadow}) is
consistent with both the single power-law extrapolation case and
a slight soft excess ($\Gamma\la 2$ for $E\la 1$ [keV]).  

 In our comparison, we use $(7.4-9.0)\times 10^{-12}$ 
$[{\rm erg\,s^{-1}\,cm^{-2}\,deg^{-2}}]$ 
as a probable range of the extragalactic 0.5-2 keV intensity, where
the  smaller value corresponds to the
single power-law form of the extragalactic component and the larger
value corresponds to the case where the 
extragalactic component steepens to a photon index of $\Gamma=2.3$ at 
$E\la 1$ [keV]. This range can be compared with the integrated intensity 
expected from the models.

\begin{figure}[tbp]
\resizebox{\hsize}{!}{\includegraphics{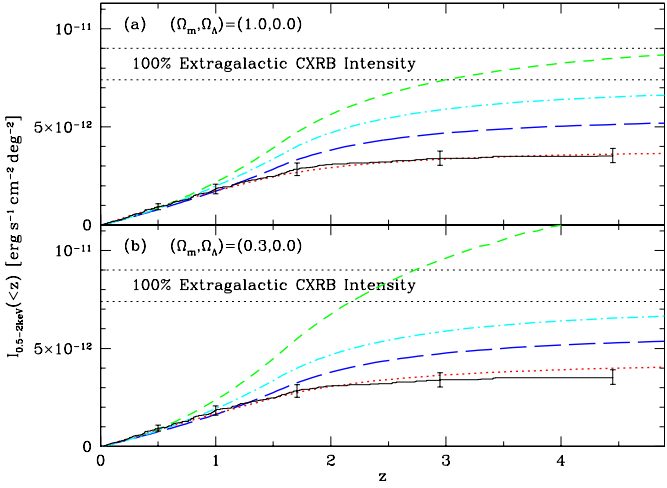}}
\caption[]{The cumulative 0.5-2 keV intensities $I(<z)$ 
are plotted as a function of redshift for the PLE, PDE,
LDDE1, and LDDE2 models for two different cosmologies as
labeled.  See caption for Fig. \ref{fig:xlfm} for 
line styles correponding to these four models. These curves
include expected contribution from sources fainter than the 
survey limit using the model extrapolations.
As a reference, the cumulative intensity $I(<z)$ of the AGNs
in the sample (see text) is also plotted (thin solid line with 
90\% errors) on each panel. This curve
represents the contribution of actually resolved and identified 
AGNs. Also the range of the 0.5-2 keV extragalactic background 
intensity (see text) is shown by two horizontal 
thin dotted lines.}
\label{fig:z_int}
\end{figure}

 In Fig. \ref{fig:z_int}, we plot the cumulative soft
X-ray (0.5-2 [keV]) intensities
of the model AGN populations as functions of redshift,
$I_{\rm 0.5-2 keV}(<z)$. 
As a reference, we have also plotted the cumulative contribution 
of the resolved AGNs in the sample, estimated by 
$\sum_{z_i<z} S_{{\rm x}i}/A(S_{{\rm x}i})$, where 
$S_{{\rm x}i}$ is the flux of the object $i$ and $A(S_{{\rm x}i})$ is 
the available survey area at this flux (Fig. \ref{fig:area}).
The portion of the model curves above this line represents
extrapolations to fainter fluxes than the limit of the deepest
survey.

 It is apparent from Fig. \ref{fig:z_int} that the PDE model
produces almost 100\% of the upper-estimate of the
CXRB intensity, giving no room for, e.g. 10\%
contribution from clusters of galaxies (M99b), in the 
$(\Omega_{\rm m},\Omega_\Lambda)=(1.0,0.0)$ universe. In the
low density universe with $(\Omega_{\rm m},\Omega_\Lambda)=(0.3,0.0)$,
the PDE model certainly overproduces the CXRB intensity. The PLE model
produces about $\sim 50\%$ of the lower estimate of the CXRB
in both cosmologies. The LDDE1 model, which best describes the data
in the observed regime, explains about $70\%$ of the
lower estimate of the CXRB intensity. The estimates 
are highly dependent on how one extrapolates the SXLF to fluxes fainter
than the survey limit. In view of this, we explore 
an alternative LDDE model, which has been adjusted to 
make $\approx 90\%$ of the lower estimate of the extragalactic CXRB 
intensity, allowing $\sim 10\%$ contribution from clusters
of galaxies. This version of the LDDE model (designated as LDDE2) has a 
fixed minimum evolution index $p_{\rm min}$ in the LDDE formula. 
Then the first case of Eq. (\ref{eq:lddez}) is replaced by:
   
\begin{eqnarray}
  e(z,L_{\rm x})\;=\; & (1+z)^{\max(p_{\rm min},{p1}-{\alpha}({\rm Log}\; 
	  {L_{\rm a}} - {\rm Log}\;L_{\rm x}))},\;\nonumber \\ 
           & (z\leq{z_{\rm c}};L_{\rm x}<L_{\rm a}). 
\label{eq:ldde2z}
\end{eqnarray}

 We do not intend to represent a particular physical picture 
behind this formula. We rather intend to search for  a formally 
simple expression 
which makes 90\% of the CXRB and is still consistent with our sample in 
the regime it covers.  We have searched for models accepted by 
the KS tests by
adjusting parameters $p_{\rm min}$, $\alpha$ and $L_{\rm a}$ by 
hand and fitting by the maximum-likelihood method with respect to
other variable parameters, requiring that the models give 
an integrated intensity of $6.7\times 10^{-12}$ 
$[{\rm erg\,s^{-1}\,cm^{-2}\,deg^{-2}}]$. The parameter values of 
such LDDE2 models are listed in Table \ref{tab:ldde2_fit}.

\begin{table}
\caption[]{Best-Fit LDDE2 Parameters} 
\begin{center}
\begin{tabular}{ll}
\hline\hline
 $(\Omega_m,\Omega_\Lambda)$ & Parameters/KS probabilities \\
\hline
\\
LDDE2 & $A=(0.88 \pm .06)\times 10^{-6}$;\,$L_*=0.85^{+.38}_{-.26}$ \\
(1.0,0.0)& $\gamma_1=0.71\pm .14$;\,$\gamma_2=2.25\pm .10$;\,
	$p1=4.82^{+.34}_{-.22}$\\
     & $z_{\rm c}=1.64\pm .16$;\,$p2=0.$(fixed);\,$p_{\min}=4.0$(fixed)\\
     & $\alpha = 1.0$(fixed);\, ${\rm Log}\,L_{\rm a}=44.1$ (fixed) \\
     & \underline{$P_{\rm KS}=0.7,0.6,0.3$ (for $L$,$z$,2D)}\\
\\
LDDE2 & $A=(1.59\pm .10)\times 10^{-6}$;\,$L_*=0.58^{+.21}_{-.14}$   \\
(0.3,0.0) & $\gamma_1=0.55\pm .16$;\, $\gamma_2=2.30\pm .08$;\, $p1=5.8\pm .3$ \\
      & $z_{\rm c}=1.57\pm .12$;\,$p2=0.$(fixed);\,$p_{\min}=3.7$(fixed) \\
      & $\alpha = 2.5$ (fixed);\, ${\rm Log}\,L_{\rm a}=44.6$ (fixed) \\
      & \underline{$P_{\rm KS}=0.99,0.5,0.4$ (for $L$,$z$,2D)}\\
\\ 	
LDDE2 & $A=(1.48\pm .09)\times 10^{-6}$;\,$L_*=0.60^{+.23}_{-.16}$   \\
(0.3,0.7) & $\gamma_1=0.57\pm .17$;\, $\gamma_2=2.21\pm .08$;\,$p1=5.3\pm .3$ \\
      & $z_{\rm c}=1.59\pm .12$;\,$p2=0.$(fixed);\,$p_{\min}=3.3$(fixed) \\
      & $\alpha = 2.5$ (fixed);\, ${\rm Log}\,L_{\rm a}=44.5$ (fixed) \\
      & \underline{$P_{\rm KS}=0.7,0.8,0.4$ (for $L$,$z$,2D)}\\
	\\ 	
\hline
\end{tabular}
\end{center}
$^{\rm a}$Units -- A: [$h_{50}^3\;{\rm Mpc^{-3}}$],\,\,
 $L_*$: [$10^{44}\;h_{50}^{-2}{\rm erg\;s^{-1}}$],\,\,
  Parameter errors correspond to the 90\% confindence level.
search (see Sect. \ref{sec:stat}).
\label{tab:ldde2_fit} 
\end{table}

 By considering LDDE2, we have shown that there still is a resonable
extrapolation of the AGN SXLF which makes up most of the soft CXRB. Of course
this is not a unique solution. One may consider LDDE1 
and LDDE2 as two possible extreme cases of how the SXLF can be
extrapolated. Further implications are discussed in 
Sect. \ref{sec:disc}.

\section{Evolution of luminous QSOs}
\label{sec:hlevol}

\begin{figure}
\begin{center}
\resizebox{\hsize}{!}{\includegraphics{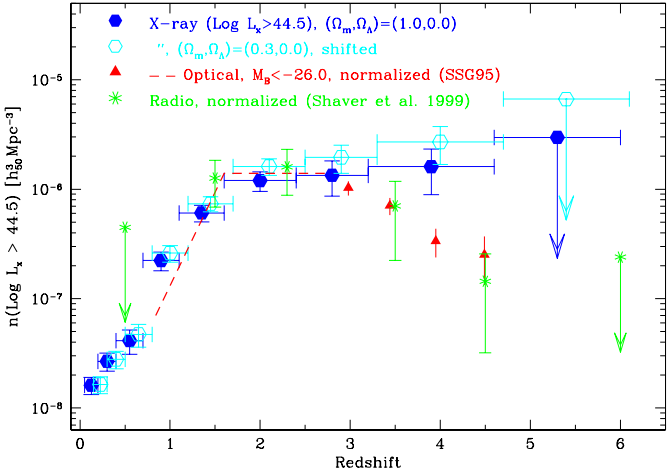}}
\end{center}
\caption[]{The comoving number density of luminous 
(${\rm Log}\;L_{\rm x}>44.5$) QSOs in our {\it ROSAT} AGN sample are plotted 
as a function of redshift for two cosmologies as labeled. 
The horizontal error bars indicate redshift bins and vertical 
error bars 1$\sigma$ errors.  The top symbol of a downward arrow 
corresponds to the 90\% (2.3 obj) upper limit. The points
for $(\Omega_{\rm m},\Omega_\Lambda)=(0.3,0.0)$ have been shifted
horizontally by $+$0.1 in $z$ for display purposes.
The numbers of the X-ray luminnous QSOs for the four highest redshift
bins are 24[32] ($1.6\leq z <2.4$), 8[12] ($2.4\leq z <3.2$),
5[7] ($3.2\leq z <4.6$), and 0[0] ($4.6\leq z <6.0$) for
$(\Omega_{\rm m},\Omega_\Lambda)=(1.0,0.0)$ [=(0.3,0.0)]. 
 For comparison, number density 
of optically-selected ($M_{\rm B}<-26$) (dashed line and filled
triangles, from SSG95) and radio-selected (stars, Shaver et al.
\cite{shav99}) QSOs, normalized to the soft X-ray selected QSO number density
at $z\sim 2.5$ are overplotted.  For the SSG95 data, this normalization
corresponds to a multiplication by a factor of 7. Shaver et al.
\cite{shav99} gave no absolute density.The optical and radio points are for
$(\Omega_{\rm m},\Omega_\Lambda)=(1.0,0.0)$.}
\label{fig:zev}
\end{figure}

 In this section, we consider QSOs with high soft X-ray 
luminosities (${\rm Log}\;L_{\rm x} > 44.5$),
where the behavior of the SXLF can be traced up to high redshifts.
Also in the high-luminosity regime, at least in the local
universe, we observe very few absorbed 
AGNs, which could cause problems with the K-correction,
in the local universe (e.g. Miyaji et al. \cite{m99b}). If this tendency
extends to the high redshift universe, our {\it ROSAT} sample 
is a good representation of luminous QSOs and the assumed 
single power-law of $\Gamma=2$ would be a reasonable one. 
Thus we here investigate the evolution of the number density
of the luminous QSOs using our sample. Fig. \ref{fig:xlf2} shows 
that the SXLF at high luminosities can be approximated by a single power-law 
($\gamma \approx 2.3$) for all redshift. Assuming this power-law with
a fixed slope, we have calculated
the number density of AGNs above this luminosity using the 
fitted normalization as described above in different
redshift bins. The results are plotted in Fig. \ref{fig:zev}
for two sets of cosmological parameters. Similar curves
for optically and radio-selected QSOs are discussed below.

 In both cases, the number density increases up to $z\sim 1.6$ and 
flattens beyond this redshift. In both cosmologies,  the number
density for $z\ga 1.7$ is consistent with no evolution. The 
Maximum-Likelihood fits in the $z\ga 1.7$, ${\rm Log}\;L_{\rm x}>44.5$
region gave density evolution indices ($\propto [1+z]^p$)
of $p=0.5\pm 2.5$ and $p=0.8\pm 2.1$ for $(\Omega_m,\Omega_\Lambda)$
=(1.0,0.0) and (0.3,0.0) respectively. Subtle differences of the density 
curves seen in Fig. \ref{fig:zev} between the two cosmologies come from 
two effects. Because different cosmologies give different
luminosity distances, some objects which do
not fall in the ${\rm Log}\; L_{\rm x}>44.5$ region for  
$(\Omega_{\rm m},\Omega_\Lambda) =$ (1.0,0.0) 
come into the sample in lower density cosmologies. Also the comoving
volume per solid angle in a certain redshift range becomes larger
in lower density cosmologies, thus the number density lowers
accordingly. These two effects work in the opposite sense and
tend to compensate with each other, but the former effect is 
somewhat stronger.   

It is interesting to compare this curve with similar ones
from surveys in other wavelengths. In Fig. \ref{fig:zev}, we overplot
number densities of 
optically- (Schmidt et al. \cite{ssg95}, hereafter SSG95) and 
radio-selected (Shaver et al. \cite{shav99}) 
QSOs for $(\Omega_{\rm m},\Omega_\Lambda) =$ (1.0,0.0). The densities
of these QSOs have been normalized to match the {\it ROSAT}-selected
QSOs at $z\sim 2.5$. This corresponds to a multiplicative
factor of 7 for the SSG95 sample. Shaver et al. (\cite{shav99}) gave
no absolute number density.  
In order to assess the statistical significance of the apparent
difference of the behavior at $z>2.7$ between the {\it ROSAT}
selected sample, we have used a Maximum-Likelihood
fit to 17 QSOs in the sample in $z\geq 2.2$ and
${\rm Log}\, L_{\rm x}\geq 44.5$. 

\begin{equation}
\frac{{\rm d}\;\Phi\,(L_{\rm x},z)}{{\rm d\;Log}\;L_{\rm x}}
\propto L_{\rm x}^{\gamma}\cdot e(z)
\end{equation}
with
\begin{equation}
e(z) = \left\{ 
	\begin{array}{ll}
	 C & (1.7 \leq z < 2.7) \\ 
	 C\, \exp[-\beta(z-2.7)] & (z > 2.7)\\
	\end{array}
       \right.,
\label{eq:ssgev}
\end{equation}
where $C$ is a constant. In above expression, $\beta = 0$ 
corresponds to no evolution even for
$z>2.7$ and $\beta= 1$ is a good description 
of the rapid decrease of optically-selected QSO number density by SSG95.  
Fig. \ref{fig:zev} shows that the radio-selected QSOs follow 
the SSG95 curve very well, but they do not have sufficient 
statistics to directly compare with the X-ray results.
We have made a Maximum-Likelihood fit with only one 
free parameter: $\beta$. Fixing $\gamma$ at 2.3, we have obtained
the best-fit value and 90\% errors (corresponding to 
$\Delta {\cal L}=2.7$) of $\beta=0.1^{+.6}_{-.5}$. The result
changed very little if we treat $\gamma$ as a free
parameter. Setting $\beta=1$ increased the ${\cal L}$ value 
by 3.3 from the best-fit value. This change in ${\cal L}$
corresponds to a 93\% confidence level. The probability that 
$\beta$ exceeds the value of 1 is $\approx 4\%$, considering 
only one side of the probability distribution.

 We have also checked statistical significance of the 
difference using the density evolution-weighted
$\langle V_{\rm e}^\prime/V_{\rm a}^\prime \rangle$ statistics,
(Avni \& Bahcall \cite{veva}), which is a variant of 
the $\langle V^\prime/V_{\rm max}^\prime \rangle$ statistics 
(Schmidt \cite{vvmax}) for the cases where surveys in different
depths are combined.
The $V_{\rm a}$ and $V_{\rm e}$ are primed to represent that
they are density evolution-weighted (comoving) volumes. If we
take $e(z)$ in Eq. (\ref{eq:ssgev}) with $\beta=1$ as the weighting function, 
$\langle V_{\rm e}^\prime/V_{\rm a}^\prime \rangle$ will give
a value of 0.5 if the sample's redshift distribution follows
the density evolution law of SSG95. An advantage of this
method over the likelihood fitting is that one can check the
consistency to an evolution law in a model-independent way, i.e.,
without assuming the shape of the luminosity function. 
Applying this statistics to 17 AGNs in  
$z\geq 2.2$, ${\rm Log}\, L_{\rm x}\geq 44.5$,
we have obtained 
$\langle V_{\rm e}^\prime/V_{\rm a}^\prime\rangle=0.65\pm 0.07$,
where the 1$\sigma$ error has been estimated by $(12 N)^{-\frac{1}{2}}$
($N$:number of objects). The same sample has given unweighted
$\langle V_{\rm e}/V_{\rm a}\rangle=0.56\pm 0.07$,
consistent with a constant number density. If we use a harder
photon index of $\Gamma=1.7$ for the K-correction, 14 objects
remain in this regime giving 
$\langle V_{\rm e}^\prime/V_{\rm a}^\prime\rangle=0.60\pm 0.08$.
  The inconsistency between the SSG95 optical results 
and our survey is thus marginal if high-redshift QSOs
have a systematically harder spectrum. 
     
\section{Discussion}
\label{sec:disc}

 In our analysis, we have found a good description of the behavior
of the SXLF from a combination of various {\it ROSAT} surveys. 
As explained in Sect. \ref{sec:kcorr}, our expression is for 
the total AGN population, including type 1 and type 2's and 
{for the observed 0.5-2 keV band, because of the uncertainties 
of contents and evolution of AGNs in various spectral classes. 
These have to be assumed to find the best-bet K-corrected AGN
evolution in the source rest frame.
A detailed discussion of this aspect is beyond the scope of this
paper.  An approach for the problem is to make 
a population synthesis modeling}, e.g.,  composed of unabsorbed 
and absorbed AGNs similar to  those of Madau et al. 
(\cite{madau94}) and Comastri et al. (\cite{coma95}) (see also
Gilli \cite{gilli99} for a recent work). If one is  
constructing a model in a similar approach using our SXLF as a 
major constraint, what the model constructor 
should do is to calculate the expected SXLFs in the {\rm observed} 0.5-2 keV 
band for all emission-line AGN populations (spectral classes) considered 
in the model (e.g. corresponding to different absorbing column
densities) and then to compare the {\em total} 
model SXLF with our LDDE1/LDDE2 expressions. One version of our own 
models constructed using this approach has been shown in M99b. 
We do not recommend the use of the expressions in Appendix A. 
as the SXLF of unabsorbed AGNs for the reasons described there and 
Sect. \ref{sec:kcorr}.   
       
 We have found two versions of LDDE expressions consistent with
our sample in the luminosity and redshift regime covered: 
one which produces $\sim 70\%$ of 
the 0.5-2 keV extragalactic CXRB (the lower etimate, see 
Sect. \ref{sec:cxrb}), and the other one which produces $\sim 90\%$,
as two relatively extreme cases on extrapolation. The real behavior 
is probably somewhere between these two. Note that we have only
calculated the contribution to the CXRB for 
${\rm Log}\;L_{\rm x}>41.7$, where fits were made.  Below
this luminosity, we observe an excess (Fig. \ref{fig:xlf2}),
which connects well with the SXLF of nearby Galaxies
(Schmidt et al. \cite{sbv}; Georgantopoulos et al. \cite{geor}),
{in the very local universe}.
This component has a local volume emissivity comparable or more to our 
sample AGNs in the 0.5-2 keV range and can contribute 
significantly to the soft CXRB. {Because of the low luminosity,
we can only detect this population in the very nearby universe
in a large-area surveys like RASS. Deep small-area surveys  
would not give enough volume to detect them, since even the deepest
part of RDS-LH can detect a ${\rm Log}\;L_{\rm x}\approx 41$ galaxy 
only up to $z\approx 0.1$.}  
  
 The X-ray emission {of this low luminosity population} is 
probably contributed by both star-formation and 
by low-activity AGNs (including LINERS). Although Georgantopoulos et al. 
(\cite{geor})'s analysis suggests that a major contribution is from
Seyfert galaxies and LINERS even at these low luminosities, 
star-formation activity can also contribute significantly to 
the X-ray emission of these low-activity AGNs 
(see Lehmann et al. \cite{leh_tru}). 
{As one extreme scenario, we assume that} the X-ray emission from these 
low-luminosity sources 
is mostly from star-formation activity and {their} volume emissivity
is assumed to evolve 
like the global star-formation rate (SFR; e.g. Madau et al. \cite{madau96}; 
Connolly et al. \cite{conn}), the integrated intensity  
would be roughly 30-40\% of the lower estimate of the CXRB intensity.
Even if the evolution of these low-luminosity sources were PLE, 
we would not detect any of them at intermediate to 
high redshifts even in the deepest 
{\it ROSAT} Survey on the Lockman Hole. 
Therefore this picture is still consistent with the
result that the {RDS-LH} did not find any
starburst galaxies.  If the above scenario is the case, the bahavior 
of the AGN component would need to be close to LDDE1 to allow room for a 
contribution from star-forming galaxies. In that case, the softer
emission from star-formation activity could contribute to the $E\la 1$[keV]
excess of the CXRB spectrum and the total extragalactic 0.5-2 [keV] 
intensity could be closer to the upper estimate.  If on the other hand, 
the large apparent local volume emissivity for the low-luminosity 
component is produced by the local overdensity and
not representative of the average present-epoch universe  
(e.g. Schmidt et al. \cite{sbv} is from a sample within 7.5 [Mpc])
and/or the X-ray evolution is {slower than the global
SFR (e.g. delayed formation of LMXB, White \& Ghosh \cite{white}), 
an LDDE2-like  behavior for the 
${\rm Log}\;L_{\rm x}\ga 41.7$ AGN component may also be possible.} 
A more detailed invetigation of 
the above scenarios and the exploration of other possibilities 
will be a topic of a future work.  

 One of the most interesting results is the evolution
of luminous QSOs discussed in Sect.\ref{sec:hlevol}. 
A comparison of the evolution and the global star-formation
rate is discussed in Franceschini et al. (\cite{franc99}),
where it is proposed that the evolution of the volume emissivity of the 
luminous QSOs evolves like the star-formation rate (SFR) of 
early-type galaxies, while that of the total AGN population (from 
the LDDE1 and LDDE2 models) may evolve like the SFR of all galaxies. 
Another interesting feature is that we find 
no evidence for a rapid decline of the QSO number density
at high redshift.
The SSG95-like decrease at $z>2.7$ is marginally rejected.
The difference may be caused by different selection criteria.
SSG95 have selected QSOs by the Ly$\alpha$ luminosity and
their QSOs are representative of more luminous QSOs 
($M_{\rm B}<-26$). Recently Wolf et al. (\cite{wolf}) reported 
a similar tendency in their sample of QSOs from one of their
CADIS fields, which typically have lower luminosities than the SSG95
sample. Our X-ray selected AGNs with ${\rm Log}\;L_{\rm x}>44.5$ have 
a seven times higher space density than SSG95 at $z\sim 2.5$ and thus 
are sampling lower luminosity QSOs than SSG95. Thus if the 
behavior of our {\it ROSAT}-selected QSOs and those of Wolf et al. 
(\cite{wolf}) is really flat, this can be indicative of different formation
epochs for lower and higher mass black holes.  
 Adding more deep {\it ROSAT} surveys would enable us to
trace the evolution in this regime with a better statistical 
significance. 
The upcoming {\it Chandra} and {\it XMM} Surveys
would extend the analysis to lower-luminsoity objects at
the highest redshifts as well as enabling us to give
spectral information to separate the K-effect and the actual
evolution of the number density.     

\section{Conclusion}

 We summarize the main conclusions of our analysis 
of the $\sim 690$ AGNs from the {\it ROSAT} surveys 
in a wide range of depths: 

\begin{enumerate}

\item Like previous works, we find a strong evolution of the
   SXLF up to $z\sim 1.5$ and a levelling-off beyond
   this redshift.   

\item We have tried to find a simple analytical description of the 
   overall SXLF. Our combined sample rejects the classical PLE model 
   with high significance. The PDE model has been marginally rejected
   statistically and also overproduces the soft CXRB.
  
\item We have found that an LDDE form (LDDE1), where the evolution rate is 
   lower at low luminosities, gives an excellent fit to the overall
   SXLF. The extrapolation of the LDDE1 form produces $\approx 60-70\%$
   of the estimated extragalactic soft CXRB.  

\item Another form of LDDE (LDDE2), which equally well 
   describes the overall SXLF from our sample, produces 
   $\approx 90\%$ of the extragalactic soft CXRB. These two LDDE 
   models may be considered as two possible extreme cases when
   one considers the origin of the soft CXRB.

\item The evolution of the number density of luminous QSOs 
   in our sample has been compared with that of optically- and 
   radio-selected QSOs. Our data are consistent with constant
   number density at $z>2.7$, while optically- and radio-selected
   QSOs show a rapid decline. The statistical significance of this
   difference is just above 2$\sigma$. Including more deep {\it ROSAT}
   surveys would trace the behavior with a better significance. 
\end{enumerate}

\acknowledgements
This works is based on a combination of extensive {\it ROSAT} surveys
from a number of groups. Our work greatly owes the effort of the {\it ROSAT}
team and the optical followup teams in producing data and the catalogs used 
in the analysis. 
In particular, we thank K. Mason, A. Schwope,
G. Zamorani, I. Appenzeller, and I. McHardy for provoding us with and 
allowing us to use their data prior to 
publication of the catalogs.  TM is supported by a fellowship from the
Max-Planck-Society during his appointment at MPE. GH acknowledges 
DLR grants FKZ 50 OR 9403 5 and FKZ 50 OR 9908 0.   

\appendix
\section{SXLF of the 'type 1' AGN sample}
\label{app:a}

 In the main part of this paper, we have concentrated on
the SXLF expression for the mixture of type 1 and type 2 AGNs for
the reasons explained in Sect.\ref{sec:kcorr}. However, since
previous works in literature mainly give expression for only
type 1 AGNs (with broad permitted lines), it is of significant
historical interest to investigate the SXLF properties for 
only type 1 AGNs. Because our samples come from several
different sources and every subsample has its own criteria for
classifying AGNs into subclasses, our expressions given here 
should not be used for any quantitative work (e.g. using
it as a starting point of a population synthesis modeling
under an assumption that they represent the {\rm unabsorbed} AGNs) 
without assessment of possible biases described in Sect.\ref{sec:kcorr}. 

 We have defined the 'type 1' AGN sample  as follows. We have included
AGNs explicitly classified  in the original catalogs as Seyfert
1-1.5's, BLRGs, and QSOs, while excluded those classified as 
Seyfert 1.8-2, NELGs, and Narrow-line Seyfert 1's (NLS1). The NLS1's 
have been excluded since they 
would not have been included in the 'broad-line' AGN samples in 
the previous works, especially  those with low-quality optical
spectra. A number of RBS objects classifed simply as 'AGNs' have been
ckecked with the NED database and/or the original spectra for
the subclassifications. For the Lockman Hole sample, we have included
objects with ID classes (a)-(c) (see Schmidt et al. \cite{rds2}) and
excluded (d)-(e). For the Marano sample, those classified as AGNs 
in  Zamorani et al \cite{marano} have been assumed to be type 1 AGNs 
unless otherwise stated, since type 2 AGNs have been explicitly noted.
The AGNs which have not been subclassified 
using the above procedure have been excluded from our 'type 1' 
sample. The fraction of AGNs included in this 'type 1' sample
are 98\% (RBS), 90\% (SA-N), 93\%(RIXOS), 88\% (NEP+Marano+UKD), and
85\% (LH) respectively.

\begin{table}
\caption[]{Best-fit Parameters for the 'type 1' Sample} 
\begin{center}
\begin{tabular}{ll}
\hline\hline
Model    & Parameters/KS probabilities\\
           $(\Omega_m,\Omega_\Lambda)$ \\
\hline
\\
PLE & $A=(4.8 \pm .3)\times 10^{-6}$;\,$L_*=0.28\pm .09$ \\
(1.0,0.0)& $\gamma_1=0.43\pm .19$;\,$\gamma_2=2.30\pm .11$;\,$p1=3.0\pm .2$\\
     & $z_{\rm c}=1.45\pm .19$;\,$p2=0.3^{+.6}_{-.8}$\\ 
     & \underline{$P_{\rm KS}=5\;10^{-4},3\; 10^{-4}, 7\; 10^{-5}$ 
	(for $L$,$z$,2D)}\\
\\
PLE & $A=(3.6 \pm .2)\times 10^{-6}$;\,$L_*=0.34\pm.10$ \\
(0.3,0.0) & $\gamma_1=0.41\pm .19$;\,$\gamma_2=2.31\pm .11$;\,$p1=3.0\pm .2$\\
     & $z_{\rm c}=1.47\pm .28$;\,$p2=0.46\pm .7$\\
     & \underline{$P_{\rm KS}=.02,.008,.002$ 
     (for $L$,$z$,2D)}\\
\\
LDDE1 & $A=(1.40 \pm .10)\times 10^{-6}$;\,$L_*=0.60^{+.32}_{-.19}$ \\
(1.0,0.0) & $\gamma_1=0.62\pm .20$;\,$\gamma_2=2.25\pm .09$;\,$p1=5.4\pm .3$\\
     & $z_{\rm c}=1.55\pm .15$;\,$p2=0.0$ (fixed) \\
     & $\alpha = 2.5\pm .8$;\, ${\rm Log} L_{\rm a}=44.2$ (fixed) \\
     & \underline{$P_{\rm KS}=0.6, 0.6, 0.6$ (for $L$,$z$,2D)};\,\\
\\ 	
LDDE1 & $A=(1.52\pm .10)\times 10^{-6}$;\,$L_*=0.55^{+.35}_{-.20}$ \\
(0.3,0.0) & $\gamma_1=0.62\pm .23$;\,$\gamma_2=2.17\pm .08$;\,$p1=5.3\pm .3$\\
     & $z_{\rm c}=1.62\pm .14$;\, $p2=0.0$ (fixed) \\
     & $\alpha = 3.0\pm .9$;\, ${\rm Log} L_{\rm a}=44.2$ (fixed) \\ 
     & \underline{$P_{\rm KS}=0.3,0.8,0.3$ (for $L$,$z$,2D)}\\
\\
\hline
\end{tabular}
\end{center}
See Captions for Tables \ref{tab:ple_pde} and \ref{tab:ldde1_fit} for
units of the parameters and other notes.
\label{tab:fit_brd} 
\end{table}

\begin{figure}
\resizebox{\hsize}{!}{\includegraphics{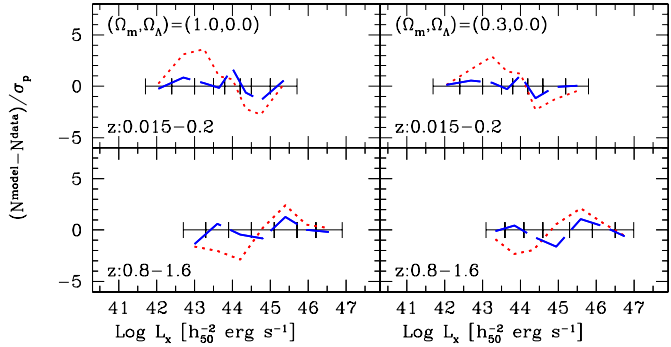}}
\caption[]{Same as Fig.~\ref{fig:nsig2} except that
this is for the 'type 1' AGN sample and that only PLE 
and LDDE1 models are plotted.}
\label{fig:nsig_br}
\end{figure}

 We have only considered the PLE and LDDE1 models in this appendix.
Table \ref{tab:fit_brd} shows best-fit parameters and KS
probabilities (see main text for details). Table \ref{tab:fit_brd} 
shows that PLE is
still rejected with a large significance for both 
cosmologies, while finding good fits with the LDDE1 form.
A plot similar to Fig.\ref{fig:nsig2} is shown for the 'type 1'
AGN sample in Fig.\ref{fig:nsig_br}. 

\end{document}